\documentclass[aps,pra,superscriptaddress,twocolumn,nofootinbib]{revtex4}

\usepackage{amstext,amsmath,amssymb,ulem}
\usepackage{color}

\usepackage[T1]{fontenc}
\usepackage[utf8]{inputenc}
\usepackage{graphicx}
\usepackage{mathptm}
\usepackage[colorlinks]{hyperref}

\newcommand{\beq}{\begin{equation}}
\newcommand{\eeq}{\end{equation}}
\newcommand{\ket} [1] {\vert #1 \rangle}
\newcommand{\bra} [1] {\langle #1 \vert}

\newcommand{\tr}{\mathop{\mathrm{tr}}}

\newcommand{\ba}{\begin{align}}
\newcommand{\ea}{\end{align}}
\newcommand{\bea}{\begin{eqnarray}}
\newcommand{\eea}{\end{eqnarray}}

\newcommand{\D}{\mathcal{D}}

\newcommand{\N}{\mathcal{N}}

\newcommand{\R}{\mathbb{R}}

\newcommand{\9}{\rangle}
\newcommand{\6}{\langle}

\newcommand{\sx}{\hat{\sigma}^x}

\newcommand{\sz}{\hat{\sigma}^z}
\newcommand{\signsymb}{\zeta}

\newcommand{\hp}{\hat{p}}
\newcommand{\hq}{\hat{q}}

\newcommand{\hZ}{\hat{Z}}
\newcommand{\hX}{\hat{X}}

\newcommand{\HCS}{\hat{H}_{\text{CS}}}
\newcommand{\HSC}{\hat{H}_{\text{SC}}}

\makeatletter
\@ifundefined{textcolor}{}
{%
 \definecolor{BLACK}{gray}{0}
 \definecolor{WHITE}{gray}{1}
 \definecolor{RED}{rgb}{1,0,0}
 \definecolor{GREEN}{rgb}{0,.4,0}
 \definecolor{BLUE}{rgb}{0,0,1}
 \definecolor{CYAN}{cmyk}{1,0,0,0}
 \definecolor{MAGENTA}{cmyk}{0,1,0,0}
 \definecolor{YELLOW}{cmyk}{0,0,1,0}
 }

\makeatother

\usepackage{bm}
\usepackage{mathtools}

\def\reals{\mathbb{R}}

\def\op#1{\hat{#1}}

\def\id{I}

\def\1{\mat{\id}}

\def\mat#1{\mathbf{#1}}

\def\controlled#1{\mathrm{C}_{#1}}
\def\CZ{{\controlled {\hat Z}}}

\newcommand{\opket}[1]{\lvert{#1}\rangle}

\newcommand{\mboxcite}[1]{\mbox{\cite{#1}}}
\renewcommand{\sout}[1]{}
\renewcommand{\mboxcite}[1]{\cite{#1}}
\begin{document} 
\title{Detecting Topological Entanglement Entropy in a Lattice of Quantum Harmonic Oscillators}
\author{Tommaso F. Demarie}
\affiliation{Centre for Engineered Quantum Systems, Department of Physics and Astronomy, Macquarie University, North Ryde, NSW 2109, Australia}
\author{Trond Linjordet}
\affiliation{Centre for Engineered Quantum Systems, Department of Physics and Astronomy, Macquarie University, North Ryde, NSW 2109, Australia}
\author{Nicolas C. Menicucci}
\affiliation{School of Physics, The University of Sydney, Sydney, NSW 2006, Australia}
\author{Gavin K. Brennen}
\affiliation{Centre for Engineered Quantum Systems, Department of Physics and Astronomy, Macquarie University, North Ryde, NSW 2109, Australia}

\begin{abstract}
The Kitaev surface-code model is the most studied example of a topologically ordered phase and typically involves four-spin interactions on a two-dimensional surface.  A universal signature of this phase is topological entanglement entropy (TEE), but due to low signal to noise, it is extremely difficult to observe in these systems, and one usually resorts to measuring anyonic statistics of excitations or non-local string operators to reveal the order. We describe a continuous-variable analog to the surface code using quantum harmonic oscillators on a two-dimensional lattice, which has the distinctive property of needing only two-body nearest-neighbor interactions for its creation.  Though such a model is gapless, it satisfies an area law and the ground state can be simply prepared by measurements on a finitely squeezed and gapped two-dimensional cluster state without topological order. Asymptotically, the continuous variable surface code TEE grows linearly with the squeezing parameter, and we show that its mixed-state generalization, the topological mutual information, is robust to some forms of state preparation error and can be detected simply using single-mode quadrature measurements. Finally, we discuss scalable implementation of these methods using optical and circuit-QED technology.
\end{abstract} 

\date{\today}
\maketitle

%\tableofcontents

%%%%%%%%%%%%%%%%%%%%%%%%%%%%%%%%%%
%\emph{Introduction.}---%
%%%%%%%%%%%%%%%%%%%%%%%%%%%%%%%%%%
%
Topological order describes a phase of matter whose correlations satisfy an area law while maintaining long-range entanglement and ground-state degeneracy impervious to all local perturbations.   These properties make such systems attractive candidates for stable quantum memories or processors \mboxcite{JP2012}.   However, the lack of a local order parameter makes measuring topological order an experimentally onerous task.  Some possibilities include measuring non-local string operators \mboxcite{LJ2008} or the statistics of anyonic excitations above the ground state, as has been demonstrated experimentally with small photonic networks \mboxcite{Lu, Pachos:07}. However, due to finite correlation lengths of local operators \mboxcite{Dusuel,LJ2008}, these methods suffer from low visibility if the system is not prepared in a pure phase with vanishing two-point correlations. 

An alternative is to study properties of the state itself that are robust to small changes in the correlation length. 
For a topologically ordered phase, the entanglement entropy of a subsystem in state $\rho_A$ is the von-Neumann entropy~$%
S(\rho_A) \equiv -\tr [\rho_A\log_2(\rho_A)]= \alpha |\partial A| - \gamma + \epsilon\,$, where $\alpha \in \R$, $|\partial A|$ is the boundary size of $A$, and $\epsilon \to 0$ for $|\partial A| \to \infty$, \mboxcite{AH2005, AK2006, ML2006}. The parameter $\gamma$ is termed the \emph{topological entanglement entropy}~(TEE)~\mboxcite{JP2012}, which is an intrinsically non-local quantity that characterizes topological phases in a variety of systems, including spin lattices such as the qubit surface code \mboxcite{Kitaev2003,AH2005}, bosonic spin liquids \mboxcite{IHM2011}, and fermionic Laughlin states \mboxcite{HZS2007}. 

While useful for numerics, actually measuring TEE in a physical system is a daunting task since extracting the von~Neumann entropy requires knowledge of the complete spectrum of the reduced state. A different option is to instead measure the Renyi entropy $S^{(\alpha)}(\rho_A)\equiv \frac{1}{1-\alpha}\log_2 \tr[\rho_A^\alpha]$ since it was shown \mboxcite{FHHW2009} that $\gamma$ is the same when replacing $S(\rho_A)$ with $S^{(\alpha)}(\rho_A)$ $\forall\, \alpha$. The value $\alpha=2$ is an attractive choice since the purity $\tr[\rho^2]$ is observable via a simple swap-test measurement on two copies of the state \mboxcite{Brennen}. A pure topological phase, such as the qudit ($d$-level spins) surface-code state~\mboxcite{Bullock}, has $\tr[\rho_A^2]=d^{1-|\partial A|}$, meaning $\gamma=\log(d)$~\mboxcite{Iblisdir}. In contrast, the purity of another area law state with no TEE, such as the qudit cluster state \mboxcite{Zhou}, is $\tr[(\rho_{A}')^{2}]=d^{-|\partial A|}$.  Thus, even using Renyi entropy one still requires a number of measurements exponential in the size $|\partial A|$ to distinguish the two phases.

In this work we study, for the first time, topological order in a continuous-variable (CV) Gaussian state~\mboxcite{gqireview} analogue of the discrete-variable surface-code state.  We describe how to prepare it efficiently using an intermediate mapping to first an ideal (infinitely squeezed) and then a physical (finitely squeezed) CV cluster state~\mboxcite{Menicucci2006, Gu2009, Menicucci2007, Menicucci2008, Flammia2009, Menicucci2011a, Aolita2011, Menicucci2013inprep}. Remarkably, the TEE of the CV surface code can be easily computed simply from quadrature measurements. We show that unlike their qubit (or qudit) counterparts, the CV surface-code state has a parent Hamiltonian that is gapless in the thermodynamic limit.  Nonetheless, the state is topologically ordered with a TEE  that asymptotically grows linearly with the squeezing parameter.  Other gapless models with topological order have been investigated in different contexts \mboxcite{FGGITT}. We propose experimental realizations for this model that are accessible with today's technology, and we conclude by analyzing the stability of the CV topological order against two forms of noise: thermalization (in the case of preparation by cooling) and noisy input states (in the case of active construction).

%%%%%%%%%%%%%%%%%%%%%%%%%%%%%%%%%%
%\emph{Ideal CV cluster states and surface-code states.}---%
%%%%%%%%%%%%%%%%%%%%%%%%%%%%%%%%%%
% 
The ideal CV cluster state \mboxcite{Zhang2006,Menicucci2006,Gu2009} is the CV analog of its qubit-based cousin \mboxcite{Raussendorf2002,Briegel2001} and may be obtained by sending zero-momentum eigenstates through pairwise controlled-$Z$ gates $\CZ_{j,k} = e^{i\hat q_j \hat q_k}$ in accord with an undirected, unweighted graph with one qumode (quantum mode---i.e., harmonic oscillator) per vertex. A CV cluster state is described completely by its \emph{nullifiers}, which are  comparable to stabilizers but with eigenvalue~0 instead of~1. Ideal CV cluster states have a complete set of nullifiers given by $\{ \hat{\eta}_j = \hp_j - \sum_{k \in \N(j)} \hq_j \}$, where $j$ runs over all vertices, and $\N(j)$ is the neighborhood of~$j$. From these nullifiers, we can construct a Hamiltonian $\HCS^{\text{ideal}} = \sum_{j} \hat{\eta}_j^2$, which has the ideal CV cluster state as its (non-normalizable) ground state. Note that $\HCS^{\text{ideal}}$ has a continuous spectrum and is therefore gapless. See \cite{Gu2009} and Appendix \ref{nosqueezingapp} for details.

\begin{figure}
\includegraphics[width=\columnwidth]{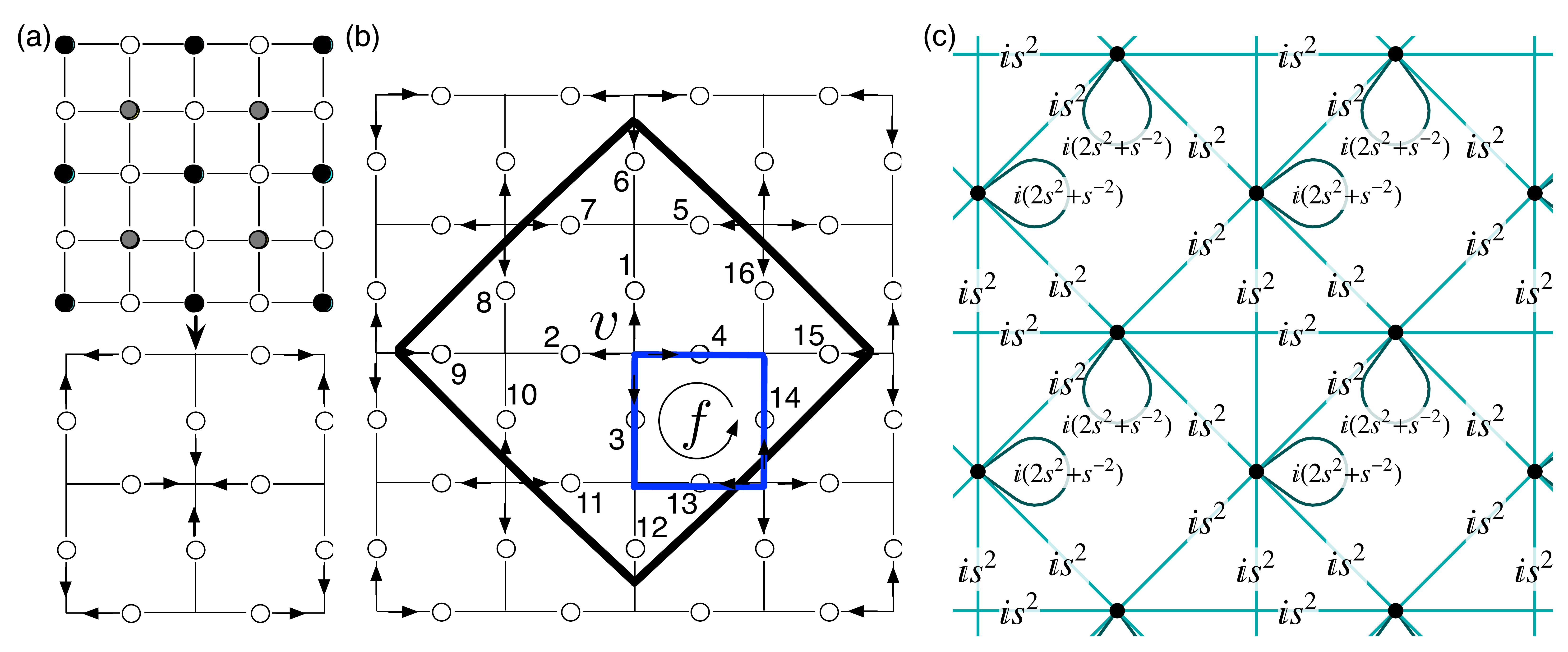}%
\caption{(a)~Measurement scheme to project the CV cluster state (top) on a square lattice into the CV surface code (bottom) described by the graph $\Lambda$. A $\hq$ measurement removes the measured (grey) node and all the links departing from it, while a $\hp$ measurement eliminates the corresponding (black) node but creates new connections among the nearest neighboring nodes.
(b)~Structure of the resultant nullifiers for the finitely squeezed surface code.  For the vertex $v$ indicated, 
$\hat{a}_v=\frac{s'}{\sqrt{8}} \Big[ \sum_{j=1}^4 \bigl( \hq_j + \frac{i}{s'^2} \hp_j \bigr) + \frac{s^2}{s'^2} \sum_{k=5}^{16} \hq_k$ \Big ] with
$s'=\sqrt{5s^2+s^{-2}}$%
, and for the face $f$, $\hat{b}_f=\frac{s}{\sqrt{8}}\bigl[(\hp_3-\hp_{13}+\hp_{14}-\hp_4) -\frac{i}{s^2} (\hq_3-\hq_{13}+\hq_{14}-\hq_4)\bigr]$.
(c)~The Gaussian pure-state graph~$\mat Z$~\mboxcite{Menicucci2011} for a section of the CV surface-code state. See main text and Appendix \ref{appZ} for details.
}
\label{scheme1}
\end{figure}

Inspired by the dynamical mapping of qubit cluster states to qubit surface codes  introduced in \mboxcite{HRD2007}, there is a simple scheme \mboxcite{JZ2008} that transforms the CV cluster state into the corresponding CV surface code. First, start with the CV cluster state, and label vertices by row and column. Then, measure in $\hp$ (in $\hq$) those vertices on rows and columns that are both odd (both even), as in Fig.~\ref{scheme1}a. This scheme is equivalent up to translation and/or inversion of the $\hp$, $\hq$ measurements.
After the measurements, we are left with a CV surface code state with a new set of nullifiers.

It is convenient to borrow notation from the qudit version of surface codes~\mboxcite{Bullock} that describes the nature of the coupling involved in the nullifiers in terms of a surface-code graph $\Lambda=\{\mathcal{V},\mathcal{E},\mathcal{F}\}$.  This graph defines an oriented surface where oscillators reside on the edges $e\in \mathcal{E}$, and the oriented edges meet at vertices $v\in\mathcal{V}$ and surround oriented faces $f\in\mathcal{F}$. Edge orientations are chosen such that for each vertex, all incident edges point toward it or all away from it (Fig.~\ref{scheme1}b). The new nullifiers are
$\hat{a}_v = \sum_{e|v\in \partial e}\hat{q}_{e}$ and $\hat{b}_f =\sum_{e\in \partial f} o(e,f)\hat{p}_{e}$, where $o(e,f)=\pm 1$ if $e$ is oriented with ($+$) or against ($-$) $f$. 
The CV surface code is the non-normalizable ground state of the quadratic Hamiltonian
$\HSC^{\text{ideal}} = \sum_v \hat{a}^{\dagger}_v\hat{a}_v + \sum_f \hat{b}^{\dagger}_f \hat{b}_f \, $, which has a fully continuous spectrum and is therefore gapless.

%------------------------------------------------------------------------------------------
%------------------------------------------------------------------------------------------

%%%%%%%%%%%%%%%%%%%%%%%%%%%%%%%%%%
%\emph{Physical CV cluster states and surface-code states.}---%
%%%%%%%%%%%%%%%%%%%%%%%%%%%%%%%%%%
%
With experimental realization in mind, we need to consider the physical case of finite squeezing. Squeezing is a Gaussian transformation, performed by the unitary operator $\hat{S}(s) = e^{-\frac{i}{2} (\log{s}) (\hq \hp + \hp \hq)}$, with $s > 0$, where $\log s$ is traditionally known as the \emph{squeezing parameter}. In the Heisenberg picture, $\hat{S}(s)^\dagger \hq \hat{S}(s)= s \hq$, and $\hat{S}(s)^\dagger \hp \hat{S}(s)= \hp/s$, such that the variance of $\hp$ (of $\hq$) after squeezing is a factor of $s^{-2}$ (of $s^{2}$) times its original value. To produce the CV cluster state in the finite-squeezing case by the canonical method~\mboxcite{Menicucci2006,Gu2009}, which is the easiest method to work with theoretically, one starts from $N$ vacuum modes $|0\9^{\otimes_N}$. Since $\hat{a}_j |0 \9^{\otimes N} = 0$ $\forall j$, the initial nullifier set is $\{ \hat{a}_j = \frac{1}{\sqrt{2}} (\hq_j + i \hp_j) \}$.
These states are then all squeezed by $\hat S(s)$, followed by the pairwise $\CZ$ couplings, yielding the transformed nullifiers
\beq
\hat{a}_j \longrightarrow \frac{s}{\sqrt{2}} \Biggl[ s^{-2} \hq_j + i \Biggl( \hp_j - \sum_{k \in \N(j)} \hq_k \Biggr) \Biggr]= \hat{\eta}_j^s \,.
\eeq
These operators satisfy the canonical commutation relations for normal-mode operators, $[ \hat{\eta}_j^s , \hat{\eta}_k^s ] = 0$ and $[ \hat{\eta}_j^s , (\hat{\eta}_k^s)^\dagger ] = \delta_{j,k}$, and therefore the CV cluster-state Hamiltonian is \mboxcite{Aolita2011}
\beq
\HCS(s) = \sum_{j=1}^N \frac{2}{s^2}\left((\hat{\eta}_j^s)^\dagger \hat{\eta}_j^s + \frac{1}{2} \right) \, .
\label{sqCSHam}
\eeq
For finite $s$, the system has a gap of~$2s^{-2}$. The prefactor provides for finite energy even in the limit of infinite squeezing where $\lim_{s\rightarrow \infty}\HCS(s)=\HCS^{\text{ideal}}$.

%%%% --------------------------------------- NEW SECTION --------------------------------------- %%%%

%\subsubsection*{Finite squeezing CV surface code}
Using the same measurement pattern as in the ideal case (Fig.~\ref{scheme1}), the finitely squeezed CV cluster state can be mapped to the finitely squeezed CV surface code.  
Taking linear combinations of neighboring cluster-state nullifiers---specifically sums of neighboring nullifiers around the $\hat{p}$-measured nodes---and alternating signed cyclic sums around $\hat{q}$-measured modes, 
one finds the general form of the surface code nullifiers ~\mboxcite{Thorn}. 
Since the finitely squeezed cluster state is Gaussian, and quadrature measurements are Gaussian operations \mboxcite{Menicucci2011}, so is the finitely squeezed surface code state.
In the case of a square lattice with toroidal boundary conditions, the nullifiers are
\begin{align}
\label{toronulli}
\hat{a}_v &=\frac{s'}{\sqrt{8}}\Biggl[\sum_{e|v\in \partial e} \left(\hat{q}_{e}+\frac{i}{{s'}^2} \hat{p}_{e} \right) +\frac{s^2}{s'^2} \sum_{\substack{v' | [v',v]\in \mathcal{E} \\ \mathclap{e |  v^\prime \in \partial e \wedge v\not\in \partial e}}} \hat{q}_e \Biggr], \nonumber \\
\hat{b}_f &=\frac{s}{\sqrt{8}}\sum_{e\in \partial f} o(e,f) \left(\hat{p}_{e}-\frac{i}{{s}^2}\hat{q}_{e} \right),
\end{align}
where $s'=\sqrt{5s^2+s^{-2}}$. See Fig.~\ref{scheme1}b.

We can construct a Hamiltonian using these nullifiers:
\begin{align}
\label{eq:HSC}
	\HSC(s) &= \sum_v \frac{8}{s'^2} \hat{a}^{\dagger}_v \hat{a}_v+\sum_f \frac{8}{s^2} \hat{b}^{\dagger}_f \hat{b}_f.
\end{align}
The squeezing dependence of the prefactors is done to ensure the Hamiltonian has finite energy for $s\to \infty$.
Unlike the discrete-variable case, this Hamiltonian is gapless in the thermodynamic limit.  This arises because the nullifiers do not define normal modes. Rather, neighboring nullifiers have nontrivial commutation relations, which allow for low-energy mode excitations.  For a square $n\times m$ ($n$ or $m$ odd) lattice with $n\leq m$ the gap is $\Delta(s)\approx 4\pi^2/s^2n^2$, and generically the system is gapless, see Appendix \ref{squeezingapp}.  Hence, in distinction to the cluster-state Hamiltonian~$\HCS(s)$, the surface-code Hamiltonian $\HSC(s)$ is gapless in the thermodynamic limit, though for infinite squeezing both models are gapless. 

%We expect that any quadratic parent Hamiltonian for a CV topologically ordered surface-code state would be gapless since the topological entanglement entropy is a continuous function of squeezing.
% A simple example is the symmetric form of the surface-code Hamiltonian with a vertex nullifier $\hat{a}_v$ that is the dual of $\hat{b}_f$ \cite{arXivvers}.

%%%%%%%%%%%%%%%%%%%%%%%%%%%%%%%%%%
%\emph{TEE for CV surface codes.---}%
%%%%%%%%%%%%%%%%%%%%%%%%%%%%%%%%%%
% 
A zero-mean $N$-mode Gaussian state is completely and uniquely described \mboxcite{Walls2008} via its symmetrized covariance matrix $\Gamma_{j,k} = \text{Re} \tr [\rho \hat{r}_j \hat{r}_k]$,
where $\hat{\bar{r}} = (\hat{q}_1, ..., \hat{q}_N, \hat{p}_1, ..., \hat{p}_N)^T$ is a $2N$-dimensional column vector of quadrature operators ~\mboxcite{Walls2008} . 
A Gaussian pure state's entanglement entropy can be calculated~\mboxcite{TD2012} in terms of the \textit{symplectic eigenvalues} of $\Gamma$
, which are the positive elements of the $N$ eigenvalue pairs $\{ \pm \sigma_j \}$ of the matrix product $i \Gamma \Omega$, with $\Omega_{j,k} = -i [ \hat{r}_j, \hat{r}_k ]$. The entropy for an $N_A$-mode Gaussian subsystem $\rho_A$ is
\begin{align}
\label{formulasetext}
S(\rho_A) = \sum_{\mathclap{\{ \sigma^A_i \}}}  \bigl[   ( \sigma^A_i + \tfrac{1}{2} ) \log_2   ( \sigma^A_i + \tfrac{1}{2} ) -  ( \sigma^A_i - \tfrac{1}{2} ) \log_2  ( \sigma^A_i - \tfrac{1}{2}  ) \bigr],
\end{align}
calculated using the reduced symplectic spectrum $\{\sigma^A_1, \dotsc, \sigma^A_{N_A}\}$ obtained deleting all of $B$'s rows and columns from the covariance matrix.

In addition to the covariance-matrix representation, every zero-mean $N$-mode Gaussian pure state can be uniquely represented by an $N$-node, undirected, complex-weighted graph whose adjacency matrix is called~$\mat Z = \mat V + i\mat U$~\mboxcite{Menicucci2011,Simon88}. When $\mat Z$ is purely imaginary (${\mat V = \mat 0}$), the state's covariance matrix becomes $\Gamma \equiv\frac{1}{2}(\Gamma_x\oplus \Gamma_p)= \tfrac 1 2 (\mat U^{-1} \oplus \mat U)$. As shown in Fig.~\ref{scheme1}c, this is the case for the CV surface-code state, for which $\mat Z = i\mat U_{\text{SC}} = i (2 s^2 + s^{-2}) \mat \id + i s^{-2} \mat A_{\text{SC}}$, where $\mat A_{\text{SC}}$ is the corresponding unweighted adjacency matrix without self-loops. From this, we see immediately that $\hp$-$\hp$ correlations (determined by $\mat U$) have range at most $1$, and $\hp$-$\hq$ correlations are zero. Further, as shown in Appendix \ref{latcorrs}, the $\hq$-$\hq$ correlations are bounded above by $\langle \hat{q}_i \hat{q}_j\rangle\leq Ce^{-(d(i,j)+1)/\xi}$ where $d(i,j)$ is the shortest distance between modes $i$ and $j$ on the surface code graph.  The constant is $C=(1+\sqrt{8s^{4}+1})^2/4(8s^2+s^{-2})$ and the correlation length $\xi$ is:
\begin{equation}
\xi\leq 2\ {\ln \Big[\frac{\sqrt{8s^4+1}+1}{\sqrt{8s^4+1}-1}\Big]}^{-1}.
\end{equation}
Therefore, because for Gaussian states all higher order correlations are generated by the linear and quadratic ones, the CV surface code state obeys an area law.  Numerically we find for squeezing $\log s=3.2$ the correlation length is $\xi=2.42$, see Fig.~\ref{corre}, and the scale factor of entropy with area is $\alpha=4.68$.
%------------------------------------------------------------------------------------------
\begin{figure}
\includegraphics[width=\columnwidth]{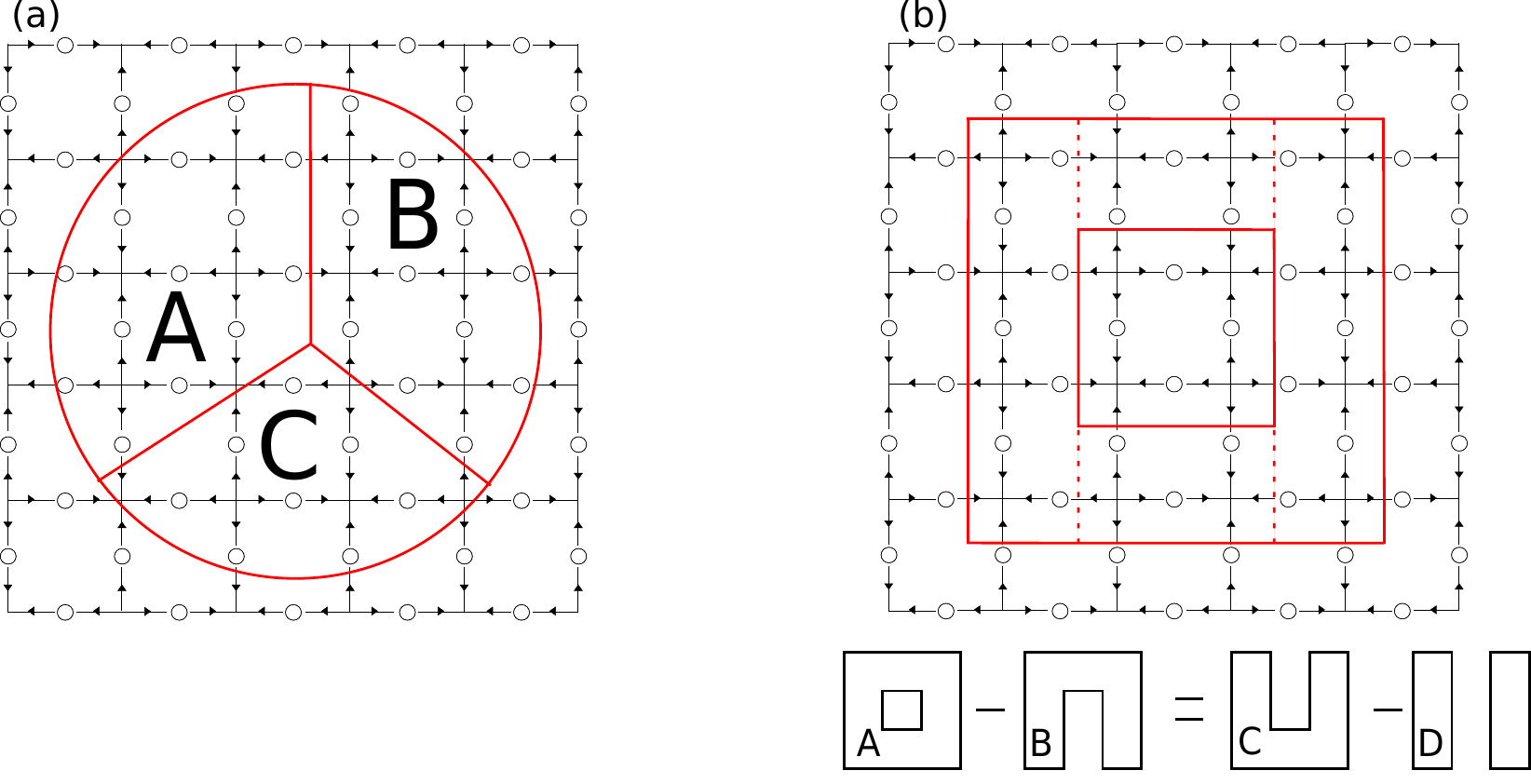}
\caption{Sections used for calculations of topological entanglement entropy by the methods of (a)~Kitaev-Preskill~\mboxcite{AK2006} and (b)~Levin-Wen~\mboxcite{ML2006}.  The areas of the regions satisfy the equality depicted.}
\label{KPfigure}
\end{figure} 
%------------------------------------------------------------------------------------------

\begin{figure}
\includegraphics[width=1\columnwidth]{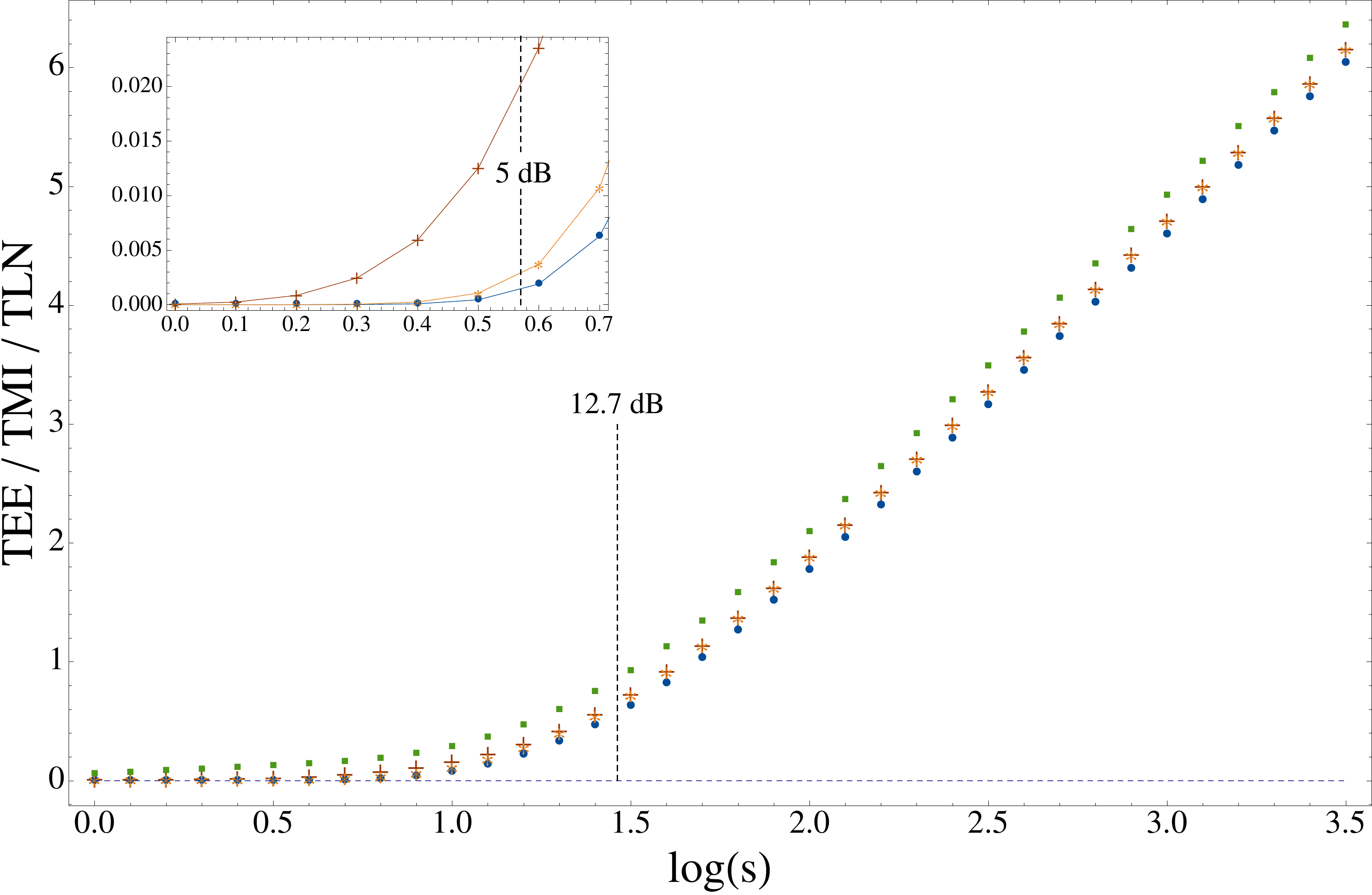}
\caption{Topological entanglement entropy (TEE) $\gamma$ as a function of the squeezing parameter $\log{s}$ for the CV surface code on a square lattice with $1296$ modes. %The continuous line represents an analytical upper bound; 
The asterisks ($\ast$) indicate the TEE calculated using Eq.~\eqref{stopoKP}; the crosses ($+$) label the TEE using Eq.~\eqref{stopoLW}; the squares ({\tiny{$\blacksquare$}}) depicts the TLN; and the circles ($\bullet $) indicate the lower bound for the topological mutual information~(TMI) $\gamma^{l}_{\rm MI}$ for noisy state preparation. The main graph shows the maximum single-mode squeezing of $12.7~$dB achieved to date~\mboxcite{Eberle2010,Mehmet2011}. Finally, the dashed line corresponds to TEE for the CV cluster state, which is always zero.   {\it Inset}: TEE/TMI for levels of multimode squeezing with 5~dB marked as achievable with current optical technology \mboxcite{Yokoyama2013}.  The TLN is above the scale shown here.  }
\label{Stopo}
\end{figure}

To study the topological properties of the CV surface-code state, we make use of two alternative (but equivalent) definitions of TEE , one introduced by Kitaev and Preskill~(KP)~\mboxcite{AK2006},
\beq
\label{stopoKP}
S_{\text{topo}}^{\text{KP}}  \equiv - ( S_A + S_B + S_C - S_{AB}- S_{BC}- S_{AC} + S_{ABC}) = \gamma \, ,
\eeq
and another by Levin and Wen (LW) \mboxcite{ML2006},
\beq
\label{stopoLW}
S_{\text{topo}}^{\text{LW}}  \equiv - \frac{1}{2} [ (S_A - S_B) - (S_C - S_D) ] = \gamma \, ,
\eeq
with regions shown in Figs.~\ref{KPfigure}a and~\ref{KPfigure}b, respectively.
If the system is not topologically ordered, these combinations are exactly zero. Thus, we say that a model is topologically ordered only when $\gamma >0$.
In our simulation we use the ground state of a $36 \times 36$ mode CV surface-code and the squeezing-dependent values of the TEE are calculated selecting from the covariance matrix the reductions corresponding to regions chosen as prescribed by Eq.~\eqref{stopoKP} and Eq.~\eqref{stopoLW}. See Fig.~\ref{Stopo} for numerical results. Further, we plot the \emph{Topological Log-Negativity} (TLN) for the KP regions based on recent results that show this quantity to be a good witness of topological order for stabiliser states of Abelian anyon models \cite{castelnovo2013, lee2013}. The log-negativity of the reduced state with support on a subsystem $A$ is \cite{vidal2002}
\begin{equation}
\mathcal{N}(\rho_A)=-\frac{1}{2} \sum_{i=1}^N \log_2[{\text{min}(1,\lambda_i(\Gamma_x \mu_A \Gamma_p \mu_A))}]\,,
\end{equation}
where $\mu_A=P_{A^{\perp}} \oplus (- P_A)$, with $P_X$ being the projector onto the modes in region $X$ and $\lambda_i$ being the $i$-th eigenvalue of the matrix argument.  Remarkably, we find that the TLN is a rather tight upper bound on the TEE with the same asymptotic slope. 
To derive this slope we consider the entanglement entropy for one mode of the smallest meaningful portion of the surface code, specifically a three-mode correlated state (see Appendix \ref{upboundTEE} for the details of the argument). The squeezing-dependent symplectic eigenvalue of the reduced covariance matrix corresponding to the mode used is given by $\sigma_1 = \frac{1}{2}(1 + 3s^4 + 2 s^8)^{1/2}(1+3s^4)^{-1/2}$. The mode entropy grows linearly with an asymptotic slope of $\lim_{s\rightarrow\infty}\frac{d\gamma(s)}{d(\log s)} = 2/\ln(2) \simeq 2.8854$, which matches the slope of TEE and TLN we find numerically for larger systems.
%We derive an upper bound to the CV TEE by considering the  entanglement entropy for one mode of the smallest meaningful portion of the surface code, specifically a three-mode correlated state \cite{arXivvers}. The upper bound for the TEE is shown in Fig.~\ref{Stopo}, with an asymptotic slope of $\lim_{s\rightarrow\infty}\frac{d\gamma(s)}{d(\log s)} = 2/\ln(2) \simeq 2.8854$.
%
%
%
%\subsubsection{Topological entropy of thermalized input states}
%
Interestingly, since the controlled-Z gates introduce additional squeezing~\mboxcite{vanLoock2007,Yoshikawa2008}, even for $s \to1$ (starting with vacuum states instead of momentum-squeezed states), the TEE is very small but non-zero.

To model noise in the state preparation, we consider a thermal state with respect to the cluster-state Hamiltonian, $\rho_{\rm CS}(\beta)=e^{-\beta \HCS(s)}/\tr[\cdot]$ as the pre-measurement initial state. This could be generated by engineering the Hamiltonian $\HCS(s)$, which is gapped for finite squeezing, and then waiting until the system reaches equilibrium with an environment at temperature $\beta^{-1}$.  Alternatively, using e.g. networks of non-interacting photons, one could start with separable modes each in a thermal input state $\rho_{\rm}(\beta)=\prod_j e^{-\beta \frac {2} {s^2} a_j^{\dagger}a_j}/\tr[\cdot]$ and then generate the thermal cluster state as before.

For such mixed states we can detect topological order by making use of the Topological Mutual Information (TMI) \mboxcite{Iblisdir}. The TMI is constructed replacing in Eq.~\eqref{stopoKP} the von~Neumann entropy $S_X$ with (half of) the mutual information $I_{X} = S_X + S_{X_c} -S_{X\cup X_c}$ between a region $X$ and its complement~$X_c$:
\beq
\label{TMI}
\gamma_{\text{MI}} \equiv - \frac 1 2 ( I_A + I_B + I_C - I_{AB}- I_{BC}- I_{AC} + I_{ABC})\, .
\eeq
%The value of the TMI lies between two bounds proposed in \mboxcite{JW2012} to avoid ambiguities in the definition \cite{arXivvers}.

Assume that, under a given sequence of symplectic transformations and homodyne detections, the ground-state covariance matrix~$\Gamma_{\text{CS}}$ of the CV cluster-state Hamiltonian~$\HCS(s)$ maps to a CV surface-code state~$\Gamma$. Because all normal modes of $\HCS(s)$ have the same frequency, the thermal covariance matrix of~$\HCS(s)$ at temperature~$\beta^{-1}$ is just $\kappa \Gamma_{\text{CS}}$ and, under the same evolution, maps to $\kappa \Gamma$, where $\kappa = \coth (\beta \epsilon_0/2)$, and $\epsilon_0$ is the energy gap of $\HCS(s)$. Note this is \emph{not} a thermal state for $\HSC(s)$ because the spectrum of $\HSC(s)$ is nonuniform.

The TMI for this class of mixed states is lower bounded by the value computed as $\kappa \to \infty$, as shown in Appendix \ref{appTEE}. This is $\gamma_{\text{MI}}^{l} = -\tfrac 1 2 \sum_{X} \signsymb(X) \sum_{i}' \log_2 (e \sigma_i^X)$, where $X$ runs over all the regions in Fig.~\ref{KPfigure}a and their complements, $\signsymb(X) = \pm 1$ in accordance with Eq.~\eqref{TMI}, and the prime on the sum indicates that we need only include the zero-temperature symplectic eigenvalues $\sigma_i^X$ for which $\sigma_i^X > 1/2$. (see Fig.~\ref{Stopo}).

%%%%%%%%%%%%%%%%%%%%%%%%%%%%%%%%%%
%%%%%%%%%%%%%%%%%%%%%%%%%%%%%%%%%%

%%%%%%%%%%%%%%%%%%%%%%%%%%%%%%%%%%
%\emph{Experimental implementations.}---%
%%%%%%%%%%%%%%%%%%%%%%%%%%%%%%%%%%
%
The CV cluster states considered in this work are canonical CV cluster states~\mboxcite{Menicucci2011}, so named because they are the states that would result if one were to use the canonical method of constructing them~\mboxcite{Zhang2006, Menicucci2006, Gu2009}. This method
%, which involves putting momentum-squeezed vacuum states through controlled-$Z$ gates of uniform weight~1, 
generates Gaussian states with graphs of the form $\mat Z = \mat V + i s^{-2}\mat \id$, where the entries of $\mat V$ are either 0 or~1. While this method is straightforward theoretically, the $\CZ$ gates are experimentally difficult and inefficient in an optical setting, where the most progress has been made~\mboxcite{vanLoock2007,Yoshikawa2008}.

More efficient and scalable optical construction methods exist~\mboxcite{vanLoock2007,Menicucci2007,Menicucci2008,Flammia2009,Menicucci2010,Menicucci2011a} but produce cluster states with uniform non-unit edge weight. 
These optical methods can produce very large states \cite{Armstrong,Treps}, including a recently demonstrated 10,000-mode cluster state with linear topology~\cite{Yokoyama2013}. Very large square lattices with toroidal~\mboxcite{Menicucci2008}, cylindrical~\mboxcite{Menicucci2008,Menicucci2011a}, and planar topology can also be made, as well as higher-dimensional lattices~\mboxcite{Wang2013}. Similar states might also be created by cooling a circuit-QED system to the ground state of $\HCS(s)$ [Eq.~\eqref{sqCSHam}]~\mboxcite{Aolita2011}. In Ref.~\mboxcite{PRTD2009} it was shown that using superconducting co-planar waveguides coupled pairwise via dissipative Cooper pair boxes, one can engineer an effective $\hq$-$\hq$ interaction between the microwave modes in neighbouring waveguides.  By changing the location of the box in the waveguides, one can also generate $\hp$-$\hp$ couplings. While the cluster-state Hamiltonian in Eq.~\eqref{sqCSHam} also has $\hq$-$\hp$ couplings, there exist parent Hamiltonians for CV cluster states (up to phase shifts) that consist only of $\hq$-$\hq$ and $\hp$-$\hp$ couplings~\mboxcite{Menicucci2011,Menicucci2013inprep}.

All of these methods can be used to efficiently produce square-lattice CV cluster states with uniform edge weight $g$, with $0<g<\tfrac{ 1}{ 4}$. In the optical case, these states can be very large (thousands of modes)~\mboxcite{Flammia2009} 
and all produce
a Gaussian state with a graph of the form~$\mat Z_g  = g \mat V + i \epsilon \mat \id$.
By squeezing each mode in $\op q$ by a factor of $\sqrt g$, $\mat Z_g \mapsto g^{-1} \mat Z_g = \mat V + i (\epsilon / g) \mat \id$. Despite being constructed by a completely different method, the resulting state is a canonical CV cluster state with effective initial squeezing $\tilde{s} = \sqrt{g/\epsilon}$. Since entanglement measures are local-unitary invariant, all of our results apply to these states if we take $s = \tilde s$. 
Furthermore, we don't need to actively perform the single-mode squeezing before measuring the TEE/TMI. We can simply rescale the outcomes of measurements on the original ($g$-weighted) state~\mboxcite{Alexander2013inprep}.

%\emph{Conclusions.}---
We have presented a model of correlated quantum harmonic oscillators in a topologically ordered continuous-variable surface-code state constructed using only Gaussian operations and with the remarkable property that its topological entanglement entropy $\gamma$ can be observed simply via quadrature measurements.  In contrast to discrete-variable systems, now $\gamma$ is a continuous function of a system parameter, specifically the squeezing; however, this is not surprising since the parent Hamiltonian for the system is gapless.  The CV surface code state can be prepared beginning in a gapped CV cluster state phase, and the topological entanglement entropy is robust to preparation errors modeled as thermal input.  This provides a practical path to observe topological order in bosonic systems using current technology.

\emph{Acknowledgments.}---This research was supported in part by the ARC Center of
Excellence in Engineered Quantum Systems (EQuS), Project No. CE110001013.  G.K.B.\ and T.F.D.\ thank the KITP where part of this work was completed with support from the National Science Foundation under Grant No. NSF PHY11-25915.  N.C.M.\ was supported by the ARC under grant No.~DE120102204. G.K.B.\ thanks M.~Hastings, Z.~Wang, R.~Ionicioiu, and S.~Rebic for helpful discussions. T.F.D.\ wishes to thank A.~Brodutch, M.~Cirio, S.~Flammia, M.~Lewenstein, S.~Singh and D.~Terno for discussions, comments and support. NCM is grateful to O.~Pfister, R.~Alexander, P.~van~Loock, S.~Flammia, A.~Doherty, and S.~Bartlett for helpful discussions. T.L.\ thanks L.~Lehman,  D.~Lombardo and V.~Siddhu for illuminating conversations.

%%%%%%%%%%%%%%%%%%%%%%%%%%%%%%%%%%
\appendix
%%%%%%%%%%%%%%%%%%%%%%%%%%%%%%%%%%

%%%%%%%%%%%%%%%%%%%%%%%%%%%%%%%%%%
\section{Infinitely squeezed CV cluster states and surface-code states}
%%%%%%%%%%%%%%%%%%%%%%%%%%%%%%%%%%
\label{nosqueezingapp}

In this appendix, we provide additional information about ideal (i.e., infinitely squeezed) CV cluster states~\mboxcite{Gu2009} and ideal CV surface-code states~\mboxcite{JZ2008}.%

%------------------------------------------------------------------------------------------
\subsubsection*{Ideal CV cluster states}
%------------------------------------------------------------------------------------------

There are many ways to construct physical CV cluster states~\mboxcite{Gu2009, Menicucci2006, vanLoock2007, Menicucci2007, Menicucci2008, Flammia2009, Aolita2011, Menicucci2011a}, all of which give slightly different states in the finitely squeezed case~\mboxcite{Menicucci2011}, Each has important differences that manifest when using them for measurement-based quantum computation~\mboxcite{Alexander2013inprep}. In the infinitely squeezed case, however, these differences become largely irrelevant, and since infinitely squeezed states are unphysical anyway, we are free to choose for their analysis the method that is simplest from a theoretical perspective. For this reason, we %will
choose the \emph{canonical method}~\mboxcite{Menicucci2006}, which is the most straightforward generalization of the qubit cluster-state preparation procedure, despite its inefficiency when used in practice~\mboxcite{vanLoock2007}.

A CV cluster state is described in terms of a continuous generalization of the Pauli group, namely, the Weyl-Heisenberg group~\mboxcite{Gu2009}. This generalization is most easily accomplished by thinking of the qubit $\sx$ and $\sz$ gates as implementing one-unit cyclic shifts in ``position'' ($|0\9 \leftrightarrow |1\9$) and ``momentum'' ($|+\9 \leftrightarrow |-\9$), respectively. The CV analogs of the qubit $\sx$ and $\sz$ gates are the translation (position-shift) operators~$\hat{X}(t)$ and boost (momentum-shift) operators~$\hat{Z}(u)$, respectively, with $t,u \in \mathbb{R}$. While for qubits the shift is by an element of $\mathbb Z_2$, in the CV case, one may implement a shift in position or momentum by any real-valued amount. These shift operators are generated by the canonical self-adjoint quadrature operators $\hp$ and $-\hq$, respectively, which satisfy $[ \hq, \hp ]=i$. Specifically,
\begin{align}
%\label{eq:}
	\sx \longrightarrow \hat{X}(t) = e^{- i t \hp}, \qquad \sz \longrightarrow \hat{Z}(u) = e^{i u \hq},
\end{align}
with the group commutator
\begin{align}
\label{eq:}
	\hat{X}(-t)\hat{Z}(-u) \hat{X}(t) \hat{Z}(u) = e^{- i t u}.
\end{align}
%Our intention now is to explain how to build a CV system that mimics the features of the qubit cluster state described previously.\\

To construct a qubit cluster state~\mboxcite{Raussendorf2002}, one starts with a collection of qubits in the $\ket +$ state ($+1$~eigenstate of the Pauli operator~$\sx$) and applies controlled-$\sz$ gates pairwise in accord with a chosen graph (traditionally, a square lattice), with one qubit per node of the graph. To construct an ideal CV cluster state by the canonical method~\mboxcite{Menicucci2006}, one begins with a collection of $N$ quantum harmonic oscillators, each prepared in a 0-eigenstate of momentum,
$|0 \9_{p}^{\otimes N}$, where the subscript means that $\hp_j |0\9_{p_j} = 0$ for each mode. A CV cluster state~$| \text{CS}\9$ is generated from these initial states through the pairwise application of controlled-Z gates $\CZ_{( j,k )}= e^{i  \hq_j \hq_k}$ upon all the nearest-neighbor modes $\6 j,k \9$ of the initial state,
\beq
\prod_{\6 j,k \9}^{N} \CZ_{( j,k )} | 0 \9_{p}^{\otimes N} = |\text{CS} \9 \, ,
\eeq
in accord with a given graph, with one mode per node of the graph.

It is worth pointing that there is some additional freedom in this construction procedure, even at the ideal level. In particular, ideal CV cluster-state graphs can have a nonzero real-valued weight~$g \in \mathbb R^*$ associated with each edge. This modifies the strength of the $\CZ$ gate represented by that edge: $\CZ_{( j,k )}[g] := e^{i g \hq_j \hq_k}$. These weights were first introduced in Ref.~\mboxcite{Menicucci2007} as a way to enable new methods of constructing CV cluster states. They have shown themselves to be very important when considering the computational properties of these states~\mboxcite{Alexander2013inprep} and when considering efficient construction of cluster states with very large graphs~\mboxcite{Flammia2009, Menicucci2011, Menicucci2011a, Menicucci2013inprep}. For the purposes of all derivations in this work, we set $g=1$, but in the main text (``Experimental Implementation''), we showed how our results apply to CV cluster states constructed with non-unit---but still uniform---weight~$g$.

Just like in the qubit case, ideal CV cluster states admit a description in terms of stabilizers. Given a stabilizer operator $\hat{K}_j$ for a state $|\psi\9$ (i.e., $\hat K_j |\psi\9 = |\psi\9$), under a unitary transformation of the state, $|\psi'\9 = \hat U |\psi\9$, $\hat K_j$ transforms as $\hat K'_j = \hat{U} \hat{K}_j \hat{U}^\dagger$ in order to preserve its status as a stabilizer: $\hat K'_j |\psi'\9 = |\psi'\9$. Note that the transformation $\hat K_j \longrightarrow \hat K'_j$ under the action of $\hat U$ is \emph{opposite} from the Heisenberg evolution of observables under the same unitary~$\hat U$. This is because we are not modeling the evolution of an observable when we evolve stabilizers. Instead, we are evolving the old stabilizer into a new stabilizer for the new state. Thus, the unitary evolution applied to the stabilizer must counteract that applied to the state in order to maintain the stabilizer's role as such.

To derive the Hamiltonian of the ideal CV cluster state, we construct the qubit case and then show the analogous steps in the CV case. Consider, for simplicity, a square-lattice graph with a qubit placed on each vertex. Each of the qubits is initially in a $|+\9$ state and hence is stabilized by a $\hat{\sigma}^x$ operator: $\hat{\sigma}^x_j |+\9_j = |+\9_j$. Next, the cluster state is created by applying a controlled-$\hat{\sigma}^z$ gate on every pair of nearest-neighbor qubits. Consequently, the stabilizers are transformed into new elements equal to
\beq
\label{qusta}
\hat{K}_j = \hat{\sigma}_j^x \prod_{k \in \N(j)} \hat{\sigma}_k^z = \hat{\sigma}_j^x \hat{\sigma}^z_N \hat{\sigma}^z_S \hat{\sigma}^z_E \hat{\sigma}^z_W\, ,
\eeq
with $\N(i)$ identifying the nearest neighbors of qubit~$j$, and with N, S, E, and W indicating the qubit to the North, South, East, and West of  qubit~$j$, respectively. Finally, the qubit cluster state is the ground state of the Hamiltonian constructed by imposing an energy penalty for violating any of the stabilizer conditions:
\beq
\hat{H}_{\text{qCS}} = - \sum_j \hat{K}_j \, . 
\eeq

To similarly define the ideal CV cluster state, we substitute the $N$ qubits on the vertices of the square-lattice graph with $N$ qumodes in the $|0\9_p$ state as defined before. The global state is therefore $| 0 \9_p^{\otimes N}$, and this is stabilized by the single-mode operators $\hat{X}_j(s)$ in the sense that $\hat{X}_j(s) | 0 \9_p^{\otimes N} = | 0 \9_p^{\otimes N}$. For CV state, there exists an equivalent way to express the stabilizer relations by using nullifiers~\mboxcite{Menicucci2011}. An operator $\hat{\eta}$ is called a nullifier for a state $|\psi\9$ when $\hat{\eta} | \psi \9 = 0$ holds. In this case, we have
\beq
\hat{X}_j(s)  | 0 \9_{p_j} = e^{- i s \hp_j}  | 0 \9_{p_j} =  | 0 \9_{p_j}\, \longleftrightarrow \hp_j  | 0 \9_{p_j} = 0 \, ,
\eeq
and the state $ | 0 \9_p^{\otimes N}$ is nullified by the set $\{ \hp_j \}$ of generators of the stabilizer group. As in the qubit case, the ideal CV cluster state $|\text{CS}\9$ is the result of the pairwise application of nearest-neighbor controlled-$Z$ gates $\CZ_{( j,k )}$ on $|0\9_p^{{\otimes N}}$. Under the $\CZ_{( j,k )}$ evolution, the quadrature operators transform as
\begin{align}
\CZ_{( j,k )} \hq_j \CZ_{( j,k )}^\dag &= \hq_j \, ,\nonumber \\
\CZ_{( j,k )} \hp_j \CZ_{( j,k )}^\dag &= \hp_j- \hq_k\, ,
\label{czsq}
\end{align}
and thus the initial state stabilizers $\{ \hat{X}_j(s) \}$ are changed into the cluster state stabilizers $\{ \hat{X}_j(s) \prod_{k \in \N(j)} \hat{Z}_k(s) \}$, which have the same form of the operators in Eq.~\eqref{qusta}. We can also express the stabilizers by~\mboxcite{Menicucci2011}
\beq
\hat{X}_j(s) \prod_{k \in \N(j)} \hat{Z}_k(s) |\text{CS} \9 = e^{- i s (\hp_j - \sum_{k \in \N(j)} \hq_k)} |\text{CS} \9 \, ,
\eeq
equivalent to the elements of the CV cluster-state nullifier set $\{ \hat{\eta}_j \}$ being defined as
\begin{equation}
 \hat{\eta}_j = \hp_j - \sum_{k \in \N(j)} \hq_k \, .
 \end{equation}
All the $ \hat{\eta}_j $ commute and are the elements of the algebra that generates the stabilizer group of $| \text{CS} \9$. Again, we can construct a Hamiltonian~$\HCS$ whose ground state is $| \text{CS} \9$ by imposing an energy penalty for violating any of the nullifier conditions:
\beq
\HCS = \sum_{j=1}^N \hat{\eta}_j^2 \, .
\eeq
Since all nullifiers~$\{\eta_j\}$ commute and have a continuous spectrum of eigenvalues~($\reals$), this Hamiltonian also has a continuous spectrum, $[0,\infty)$, and is therefore gapless.

%------------------------------------------------------------------------------------------
\subsubsection*{Ideal CV surface-code states}
%------------------------------------------------------------------------------------------

Now we present a general description of the ideal CV surface code using a representation reminiscent of case for qudits~\mboxcite{Bullock}. As stated in the main text, consider a graph $\Lambda=\{\mathcal{V},\mathcal{E},\mathcal{F}\}$ where $\mathcal{V}$ is a vertex set, $\mathcal{E}$ is an edge set, and $\mathcal{F}$ is a face set.  We assume the graph is oriented and that the faces inherit this orientation.  Each quantum mode is located on an edge $e_j\in\mathcal{E}$, with the orientation of any edge~$e$ determined by $e=[v,v']$ for the base starting at vertex $v$ and the head a vertex $v'$.

The construction for the ideal CV surface-code state in the main text focussed on creating the state by starting with an ideal CV cluster state and then performing quadrature measurements on selected modes of the system in accord with a simple pattern~\mboxcite{JZ2008} in analogy to the qubit case~\mboxcite{HRD2007}. This was involved specializing to the simple case of $\Lambda$ being a planar square lattice with edge orientations chosen such that at any vertex $v$, all incident edges point toward $v$, or all point away from $v$, and nullifiers were derived under these assumptions. This Appendix will diverge from the main text and make no such assumptions about the edge orientations.

In this more general case, the stabilizers for the surface code are
\begin{equation}
\label{eq:idealCVSCstab}
\hat{A}_{v}(t)=\prod_{e|v\in\partial e} \hat{Z}_e[ o(e,v) t], \qquad \hat{B}_{f}(s)=\prod_{e\in \partial f}\hat{X}_j[-o(e,f) s],
\end{equation}
where
\begin{align}
	 o(e,v) &=
	 \begin{cases}
		+1 & \text{if $e\in [v,\cdot]$}, \\
		-1 & \text{if $e\in [\cdot,v]$},
	\end{cases}
	\nonumber \\
	 o(e,f) &=
	 \begin{cases}
		+1 & \text{if $e$ is oriented the same as $f$}, \\
		-1 & \text{otherwise},
	\end{cases}
\end{align}
where the dot ($\cdot$) stands for any vertex. By construction, the stabilizers commute: $[\hat{A}_v,\hat{A}_{v'}]=[\hat{B}_f,\hat{B}_{f'}]=[\hat{A}_v,\hat{B}_f]=0$.  The $+1$ co-eigenspace of $\hat{A}_v$ and $\hat{B}_f$ is the surface-code subspace.

Oriented edges are useful when making the connection to qudits, but they are unnecessary in the special (and original) case of the qubit surface code~\mboxcite{Kitaev2003}. In this case, the CV stabilizers in Eq.~\eqref{eq:idealCVSCstab} function once again as the CV analogs of the surface-code stabilizers for qubits:
\begin{align}
\hat{A}_v(t) \longleftrightarrow \hat{A}_v^q &=  \prod_{j \in v} \hat{\sigma}_j^z\, , & \hat{B}_f(s) \longleftrightarrow \hat{B}_f^q &= \prod_{j \in \partial f} \hat{\sigma}_j^x \, .
\end{align}
In the special case considered in the main text---i.e., when $\Lambda$ is a planar square lattice with edge orientations chosen such that at any vertex $v$, all incident edges point toward $v$, or all point away from $v$, $o(e,v)$ is a constant ($\pm 1$) for any vertex~$v$. Thus,  it amounts to only a sign flip on $t$ in Eq.~\eqref{eq:idealCVSCstab}, which has no effect on a stabilizer's role as such, and in the main text we ignore it for simplicity. Under these conditions, the stabilizers in Eq.~\eqref{eq:idealCVSCstab} become
\begin{equation}
\hat{A}_{v}(t) \longrightarrow e^{i \hat{a}_v t},\qquad \hat{B}_f(s) = e^{i \hat{b}_f s}  \, ,
\end{equation}
where, as noted in the main text, the nullifiers $\hat{a}_v$, $\hat{b}_f$ are
\begin{align}
\hat{a}_v &= \sum_{e|v\in \partial e}\hat{q}_{e}\, , \\
\hat{b}_f &= \sum_{e\in \partial f} o(e,f)\hat{p}_{e}\, .
\end{align}
If, on the boundary, one of the edges is missing, then that mode is not included in the nullifiers. 
The Hamiltonian whose ground state is the ideal CV surface code is given by
\beq
\hat{H} = \sum_v \hat{a}^{\dagger}_v\hat{a}_v + \sum_f \hat{b}^{\dagger}_f \hat{b}_f \, .
\eeq
While we have written this Hamiltonian as if it were that of a system of coupled oscillators, this connection is spurious in the ideal case [although it will be correct in the finitely squeezed case; see Eq.~\eqref{HSC}]. Since the ``mode operators''~$\hat a_v$ and $\hat b_f$ are actually (Hermitian) quadrature operators that all commute, this Hamiltonian has a continuous spectrum~$[0,\infty)$ and is therefore gapless for any number of systems.

%%%%%%%%%%%%%%%%%%%%%%%%%%%%%%%%%%
\section{Hamiltonian of the finitely squeezed CV surface~code}
%%%%%%%%%%%%%%%%%%%%%%%%%%%%%%%%%%
\label{squeezingapp}

In this section we provide details on the form of the finitely squeezed CV surface-code Hamiltonian and its spectrum. The explicit form of the surface-code nullifiers for a square lattice (while allowing for smooth or rough open boundaries) are obtained by taking linear combination of neighboring nullifiers of the finitely squeezed CV cluster state. Given a set of exact nullifiers for a Gaussian state~\mboxcite{Menicucci2011}, one can obtain new nullifiers by a three-step process:
\begin{enumerate}
\item Given a quadrature measurement $\hat x_j$ to be made on mode~$j$, where $\hat x \in \{\hat q, \hat p\}$, using linear combinations of the original nullifiers, write a new set of nullifiers such that the canonically conjugate local quadrature $\hat y_j$ (where $[\hat x_j, \hat y_j] = \pm i$) appears in only one nullifier in the new set.
\item In each new nullifier, replace $\hat x_j$ with the real-valued measurement outcome.
\item Eliminate the nullifier that contains $\hat y_j$.
\end{enumerate}
We always assume that the outcome of the measurement is~0 because any other outcome would merely result in the same state up to displacements in phase space. These displacements can always be undone by local unitaries and therefore do not change any entanglement measure we might want to calculate.

To obtain the face nullifiers $\hat{b}_f$, one sums the cluster-state nullifiers immediately adjacent to the node in question (along the cardinal directions) with orientation-dependent signs (e.g., $\hat{\eta}_{N}^s+\hat{\eta}_{S}^s-\hat{\eta}_{E}^s-\hat{\eta}_{W}^s$). The vertex nullifiers~$\hat{a}_v$ are more complex and require next-nearest-neighbor nullifiers to be added to the sum in order to achieve step~1 in the procedure above. The result for a lattice with possibly incomplete vertices and faces is
\begin{align}
%\label{eq:}
	\hat{a}_v &= \frac{s_v}{\sqrt{2V(v)(1+(s/s_v)^2)}}\nonumber \\
	&\qquad \times \Biggl[\sum_{e|v\in \partial e}%o(e,v)
\left(\hat{q}_{e}+\frac{i}{{s_v}^2} \hat{p}_{e} \right) + \frac{s^2}{s_v^2}\sum_{v'|[v',v]\in \mathcal{E}}\sum_{e| v'\in \partial e}  \hq_e \Biggr] \nonumber \\
	\hat{b}_f &=\frac{s}{\sqrt{2|\partial f|}}\sum_{e\in \partial f}o(e,f) \left(\hat{p}_{e}-\frac{i}{{s}^2}\hat{q}_{e}\right),
\end{align}
%with $s_v=\sqrt{\frac{1+V(v)s^4}{s^2}}$
with $s_v=\sqrt{V(v)s^2+s^{-2}}$, $V(v)$ valence of vertex $v$ and $|\partial f|$ boundary size of the lattice face. 

Note that now there is a dependence on the lattice position of the vertex or plaquette considered. 
The nullifiers satisfy the following commutation relations:
\begin{align}
\label{comms}
	[\hat{a}_v,\hat{a}^{\dagger}_{v'}] &= 
	\begin{cases}
		1 & \text{if $d(v,v')=0$,} \\
		\frac{(s_v^2+s_{v'}^2+s^2(V(v)+V(v')))/2s_vs_{v'}}{[V(v)V(v')(1+(s/s_{v})^2)(1+(s/s_{v'})^2)]^{1/2}} & \text{if $d(v,v')=1$,} \\
		\frac{2s^2/s_vs_{v'}}{[V(v)V(v')(1+(s/s_{v})^2)(1+(s/s_{v'})^2)]^{1/2}} & \text{if $d(v,v')=\sqrt{2}$,} \\
		\frac{s^2/s_vs_{v'}}{[V(v)V(v')(1+(s/s_{v})^2)(1+(s/s_{v'})^2)]^{1/2}} & \text{if $d(v,v')=2$,} \\
	0 & \text{if $d(v,v')>2$,}
	\end{cases}
	\nonumber \displaybreak[0] \\
	[\hat{b}_f,\hat{b}^{\dagger}_{f'}] &=
	\begin{cases}
		1 & \text{if $f=f'$,} \\
		\frac{1}{\sqrt{|\partial{f}||\partial{f'}|}} & \text{if $[f,f'] \in \mathcal{E}$,} \\
		0 & \text{otherwise,}
	\end{cases}
	\nonumber \displaybreak[0]  \\
	[\hat{b}_f,\hat{b}_{f'}] &= [\hat{a}_v,\hat{a}_{v'}] = [\hat{a}_v,\hat{b}_{f}] = [\hat{a}_v,\hat{b}^{\dagger}_{f}] = 0.
\end{align}
Here $d(v,v')$ is the Euclidean distance between the two vertices where the edge lengths of the graph are unit length. The Hamiltonian is given by
\begin{align}
\label{HSC}
	\HSC(s)&=\sum_v \frac{2V(v)(1+s^2/s_v^2)}{{s_v}^2} \hat{a}^{\dagger}_v \hat{a}_v+\sum_f \frac{2 |\partial f|}{s^2} \hat{b}^{\dagger}_f \hat{b}_f,
\end{align}
and the squeezing dependence of the prefactors for $\hat{h}_V$ and $\hat{h}_F$ ensures the Hamiltonian has finite energy in the infinitely squeezed limit:
\begin{equation}
\label{finenlimit}
\lim_{s\rightarrow \infty} \HSC(s) =\sum_v \Biggl(\sum_{e|v\in \partial e} %o(e,v)
\hat{q}_{e}\Biggr)^2+\sum_f \Biggl(\sum_{e\in \partial f} o(e,f)\hat{p}_{e}\Biggr)^2\, .
\end{equation}
Here we have used the fact that in the infinite-squeezing limit, each vertex nullifier involves a sum of $\hq$'s around that vertex and its four neighboring vertices, and since they all commute, the parent Hamiltonian is simply the squared sum of $\hq$'s around each vertex.

Now we compute the gap of the surface code Hamiltonian, Eq.~\eqref{HSC}.   We first consider a square $n\times m$ lattice with toroidal boundary conditions:
\begin{align}
%\label{eq:}
	[\hat{a}_v,\hat{a}^{\dagger}_{v'}] &= w[d(v,v')], \nonumber \\
	[\hat{b}_f,\hat{b}^{\dagger}_{f'}] &=x[d(f,f')], \nonumber \\
	[\hat{a}_v,\hat{a}_{v'}] &= [\hat{b}_f,\hat{b}_{f'}] = [\hat{a}_v,\hat{b}_{f}] = [\hat{a}_v,\hat{b}^{\dagger}_{f}] = 0,
\end{align}
where $d(v,v')$ (respectively, $d(f,f')$) is the Euclidean distance between vertices (faces) on the unit-edge-length lattice (dual lattice): 
\begin{align}
%\label{eq:}
	w(0) &=1, \quad w(1)=\frac{(1+ 8s^4)}{4(1+5s^4)}, \quad w(\sqrt{2})= \frac{s^4}{2(1+5s^4)}, \nonumber \\
	w(2) &= \frac{s^4}{4(1+5s^4)}, \qquad w(d>2)=0,
\end{align}
and
\begin{align}
	 x(0)=1, \quad x(1)=\frac{1}{4}, \quad x(d>1)=0. 
\end{align}
On a torus $|\mathcal{E}|=2nm$,$|\mathcal{F}|=nm$, and $|\mathcal{V}|=nm$.  We first focus on the case where $n\times m$ is odd such that there are $|\mathcal{E}|$ independent nullifiers which therefore span the space of all the physical-mode annihilation operators.  Introducing normal-mode operators
\begin{align}
%\label{eq:}
	\hat{c}_j&=\sum_{r=0}^{n-1}\sum_{s=0}^{m-1} \alpha^{(j)}_{r,s} \hat{a}_{v_{r,s}}, & \hat{d}_j&=\sum_{r=0}^{n-1}\sum_{s=0}^{m-1}\beta^{(j)}_{r,s} \hat{b}_{f_{r,s}},
	\label{normalmodeops}
\end{align}
where the vertices at the lattice sites have vertex coordinates $\{v_{r,s}\}$ and the sites of the dual lattice have face coordinates $\{f_{k,l}\}$,
the Hamiltonian is
\begin{equation}
\HSC(s)=\sum_{j}\frac{8\omega_j }{{s'}^2}\hat{c}^{\dagger}_j \hat{c}_j + \sum_{j} \frac{8\delta_j }{s^2} \hat{d}^{\dagger}_j \hat{d}_j.
\end{equation}
To find the normal-mode frequencies, we need to solve the equations
\begin{align}
%\label{eq:}
	\left[\hat{c}_j,\sum_v \hat{a}^{\dagger}_v \hat{a}_v \right] &= \omega_j \hat{c}_j, & \left[\hat{d}_j,\sum_f  \hat{b}^{\dagger}_f\hat{b}_f \right] &= \delta_j \hat{d}_j.
\end{align}
These two linear equations can be vectorized and rewritten as
\begin{equation}
M_v \opket{\alpha^{(j)}}= \omega_j \opket{\alpha^{(j)}},\quad M_f \opket{\beta^{(j)}}= \delta_j \opket{\beta^{(j)}} \, ,
\label{eqsnormals}
\end{equation}
where $\opket{\alpha^{(j)}}$ is the vectorized form of the operator~$\hat c_j$, and $\opket{\alpha^{(j)}}$ is that for $\hat d_j$. Defining the shift operator $\hat X_r=\sum_{k=0}^{r-1} {\ket{k\oplus_r 1}\bra{k}}$, we have
\begin{align}
%\label{eq:}
	M_v &=\hat\id_{nm}+w(1)\bigl[\hat\id_{n}\otimes (\hat X_m+\hat X_m^{\dagger})+(\hat X_n+\hat X_n^{\dagger})\otimes \hat\id_m\bigr] \nonumber \\
	&\quad+w(\sqrt{2})\bigl[\hat X_n\otimes \hat X_m+\hat X_n^{\dagger}\otimes \hat X_m^{\dagger}+\hat X_n\otimes \hat X_m^{\dagger}+\hat X_n^{\dagger}\otimes \hat X_m\bigr] \nonumber \\
	&\quad+w(2)\bigl[\hat\id_{n}\otimes (\hat X^2_m+\hat X_m^{2\dagger})+(\hat X^2_n+\hat X_n^{2\dagger})\otimes \hat\id_m\bigr], \nonumber \\
	M_f &=\hat\id_{nm}+x(1)\bigl[\hat\id_{n}\otimes (\hat X_m+\hat X_m^{\dagger})+(\hat X_n+\hat X_n^{\dagger})\otimes \hat\id_m\bigr].
\end{align}
The linear equations Eq.~\eqref{eqsnormals} can be solved in the Fourier basis via $\hat F_n\otimes \hat F_m$ where $\hat F_r=\frac{1}{\sqrt{r}}\sum_{j,k=0}^{r-1}e^{i2\pi jk/r}\ket{j}\bra{k}$.
\begin{align}
%\label{eq:}
	&\{\omega_{j}\} =\left\{1+2w(1)\left[\cos\left(\frac{2\pi j_x}{n}\right)+\cos\left(\frac{2\pi j_y}{m}\right)\right] \right. \nonumber \\
	&\qquad +
2w(\sqrt{2})\left[\cos\left(\frac{2\pi j_x}{n}+\frac{2\pi j_y}{m}\right)+\cos\left(\frac{2\pi j_x}{n}-\frac{2\pi j_y}{m}\right)\right] \nonumber \\
	&\qquad +
\left. 2w(2)\left[\cos\left(\frac{4\pi j_x}{n}\right)+\cos\left(\frac{4\pi j_y}{m}\right)\right]\right\}_{j_x=0, j_y=0}^{n-1, m-1}, \nonumber \\
	&\{\delta_{j}\} = \left\{1+2x(1)\left[\cos\left(\frac{2\pi j_x}{n}\right)+\cos\left(\frac{2\pi j_y}{m}\right)\right]\right\}_{j_x=0,j_y=0}^{n-1,m-1}, \nonumber \\
	&\opket{\beta^{(j)}}=\opket{\alpha^{(j)}} =\hat F_n \ket{j_x}\otimes \hat F_m\ket{j_y},
\end{align}
treating $j=(j_x,j_y)\in\mathbb{Z}_n\times \mathbb{Z}_m$ as a collective index.  The gap energy is the lowest-frequency mode energy:
\begin{align}
%\label{eq:}
	\Delta(s)={\rm min}_{j_x,j_y} \left\{\frac {8s^2\omega_j}{1+5s^4}, \frac {8\delta_j}{s^2} \right\}.
\end{align}
For large systems sizes, $n,m\gg 1$, and choosing without loss of generality $n\leq m$, the gap is
 \begin{equation}
\Delta(s)\approx \frac {4\pi^2}{s^2n^2}.
\label{gap}
\end{equation}
If $n, m$ even on the torus, then not all face nullifiers are independent (simply bicolor the faces, and assign plus signs to face operators on one color and minus signs on the other, and then add to get zero).  Thus, the Hamiltonian $\HSC(s)$ is underconstrained, and there is an exact gapless zero mode.  For a square lattice with planar boundaries, there are boundary effects, but this makes only a small modification to the gap, which still scales like the inverse of the system size for large lattices.

%%%%%%%%%%%%%%%%%%%%%%%%%%%%%%%%%%
\section{Topological S-matrix for symmetric CV surface code}
%%%%%%%%%%%%%%%%%%%%%%%%%%%%%%%%%%
\label{Smat}

One of the defining characteristics of topologically ordered matter is the existence of non-local interactions between particle-like excitations, which are described in terms of a scattering matrix.  The scattering matrix can be evaluated in terms of expectation values of products of ribbon operators around a torus \mboxcite{JP2012}.   We show how these arise in the context of continuous-variable topologically ordered states.

For simplicity, consider the symmetric Hamiltonian mentioned in the text:
\begin{align}
%\label{eq:}
	\HSC'(s)= \frac{8}{s^2}\sum_v \hat{a}'^{\dagger}_v\hat{a}'_v+\frac{8}{s^2}\sum_f \hat{b}^{\dagger}_f\hat{b}_f ,
\end{align}
where 
\begin{align}
%\label{eq:}
	\hat{b}_f&=\frac{s}{\sqrt{8}}\sum_{e\in \partial f} o(e,f) \left(\hat{p}_{e}-\frac{i}{{s}^2}\hat{q}_{e} \right), \nonumber \\
	\hat{a}'_v&=\frac{s}{\sqrt{8}}\sum_{e|v\in \partial e} \left(\hat{q}_{e}+\frac{i}{{s}^2} \hat{p}_{e} \right).
\end{align}
For the case $n, m$ even, because of the simple structure of the vertex and face nullifiers, there are only $n-1$ independent vertex and $m-1$ independent face nullifiers, so in fact these operators do not span the space of physical modes. Rather, there are two gapless modes.  To see this, repeat the argument above of bicoloring the lattice (respectively, the dual lattice) and assigning $+1$ weight to vertex (face) nullifiers of one color, assigning$-1$ to the other, and adding to get zero. 

On the torus, there are non-local string symmetries of the ground-state subspace:
\begin{align}
%\label{eq:}
	\hZ_{\mathcal{P}}(t) &=\exp \left[it\sum_{e\in \mathcal{P}}\frac{o(e)}{\sqrt{1-s^{-4}}}\left(\hp_e-\frac{i}{s^2}\hq_e \right)\right], \nonumber \\
	\hX_{\tilde{\mathcal{P}}}(r) &=\exp\left[-ir\sum_{e\in \tilde{\mathcal{P}}}\frac{f(e)}{\sqrt{1-s^{-4}}}\left(\hq_e+\frac{i}{s^2}\hp_e\right)\right],
\end{align}
where $\mathcal{P}$ and $\tilde{\mathcal{P}}$ are oriented paths on the lattice and dual lattice, respectively.  Here $o(e)=\pm 1$ if the edge $e$ is oriented in the same (opposite) direction as $\mathcal{P}$, and $f(e)=\pm 1$ if the edge $e$ has the same (opposite) framing to the path $\tilde{\mathcal{P}}$.  The framing of the path $\tilde{\mathcal{P}}$ is to the right, normal to its direction.    Since each string touches an even number of modes of each nullifier with opposite signs due to edge orientations, then we have by construction:
\begin{align}
%\label{eq:}
	[ \hZ_{\mathcal{P}}(t),\hat{a}'_v]&= [ \hZ_{\mathcal{P}}(t),\hat{a}'^{\dagger}_v]=[ \hZ_{\mathcal{P}}(t),\hat{b}_f]=0,\nonumber \\
	[ \hX_{\tilde{\mathcal{P}}}(r),\hat{b}_f]&=[ \hX_{\tilde{\mathcal{P}}}(r),\hat{b}^{\dagger}_f]=[ \hX_{\tilde{\mathcal{P}}}(r),\hat{a}'_v]=0.
\end{align}
However, $[ \hZ_{\mathcal{P}}(t),\hat{b}^{\dagger}_f]\neq 0$, and $[ \hX_{\tilde{\mathcal{P}}}(t),\hat{a}'^{\dagger}_v]\neq 0$, so these string operators are not symmetries of the Hamiltonian $\HSC'(s)$.  However they are symmetries of the ground subspace $\mathcal{H}'_{GS}$.  To see this notice that  the normal mode operators for this symmetric model are defined as in Eq. \eqref{normalmodeops}, call them $\hat{c}'_j$ and $\hat{d}_j$.   Since $\hat{c}'_j$ and $\hat{d}_j$ are linear combinations of the nullifiers $\hat{a}'_v$ and $\hat{b}_f$ operators respectively, we have
\begin{align}
%\label{eq:}
	\hat{c}'_j[\hZ_{\mathcal{P}}(t)\ket{GS}]&=\hZ_{\mathcal{P}}(t)\hat{c}'_j\ket{GS}=0, \nonumber \\
	\hat{d}_j[\hZ_{\mathcal{P}}(t)\ket{GS}]&=\hZ_{\mathcal{P}}(t)\hat{d}_j\ket{GS}=0, \nonumber \\
	\hat{c}'_j[\hX_{\tilde{\mathcal{P}}}(r)\ket{GS}]&=\hX_{\tilde{\mathcal{P}}}(r)\hat{c}'_j\ket{GS}=0, \nonumber \\
	\hat{d}_j[\hX_{\tilde{\mathcal{P}}}(r)\ket{GS}]&=\hX_{\tilde{\mathcal{P}}}(r)\hat{d}_j\ket{GS}=0.
\end{align}
For contractable paths $\mathcal{P}$ and $\tilde{\mathcal{P}}$, the generators of $\hZ_{\mathcal{P}}(t)$ and $\hX_{\tilde{\mathcal{P}}}(r)$ are linear combinations of $\hat{b}_f$ or $\hat{a}'_v$ operators inside the loops.  

We can compute the topological S-matrix for this model by computing the monondromy~\mboxcite{Nayak2008}. We loosely refer to the string operators as worldlines for particles that are created out of the vacuum, make an excursion around one non-contractible loop on the torus, and then annihilate.   There are two types of particles:  electric charges with charge $t$ associated with loops $\hZ_{\mathcal{P}}(t)$ and magnetic fluxes with flux $r$ associated with loops $\hX_{\tilde{\mathcal{P}}}(r)$.  The scattering-matrix element associated with a $t$ electric charge braiding around an $r$ magnetic flux is
\begin{equation}
S_{r,t}=\bra{GS}\hZ^{-1}_{\mathcal{P}_1}(t) \hX^{-1}_{\tilde{\mathcal{P}_1}}(r)\hZ_{\mathcal{P}_2}(t) \hX_{\tilde{\mathcal{P}_2}}(r)\ket{GS},
\end{equation}
where $\mathcal{P}_{1,2}$ are two loops along the vertical direction, and $\tilde{\mathcal{P}}_{1,2}$ are two loops along the horizontal direction.  This can be evaluated by using the symmetry of the loop operators:
\begin{align}
%\label{eq:}
	S_{r,t}&=\bra{GS}\hZ_{\mathcal{P}_1}(t)\hZ_{\mathcal{P}_1}(-t) \hX_{\tilde{\mathcal{P}_1}}(r) \hX_{\tilde{\mathcal{P}_2}}(-r)\hZ_{\mathcal{P}_1}(-t) \hX_{\tilde{\mathcal{P}_1}}(-r) \nonumber \\
	&\qquad \times \hZ_{\mathcal{P}_2}(t) \hX_{\tilde{\mathcal{P}_2}}(r)\ket{GS} \nonumber \\
	&=e^{i rt},
\end{align}
which is the expected statistics based on the group commutator for the Weyl representation of the Heisenberg group:
\begin{align}
%\label{eq:}
	e^{-i t q}e^{i r p}e^{itq}e^{-irp}=e^{irt}.
\end{align}
Here we have used the facts that $\hZ^{-1}_{\mathcal{P}}(t)=\hZ_{\mathcal{P}}(-t)$ and $\hX^{-1}
_{\tilde{\mathcal{P}}}(r)=\hX_{\tilde{\mathcal{P}}}(-r)$ and that the only nontrivial action is at four intersection mode locations $\{e_a,e_b,e_c,e_d\}$.  Note there is a factor $(-1)^{f(e_a)_f(e_c)+f(e_d)+o(e_a)+o(e_c)+o(e_d)}$ also in the exponent which we have assumed is equal to one.  If for our chosen edge orientations this is $-1$, we simply redefine either $r\rightarrow -r$ or $t\rightarrow -t$. Note that in the limit of infinite squeezing, $s \to \infty$, the string operators become unitary.

%%%%%%%%%%%%%%%%%%%%%%%%%%%%%%%%%%
\section{General properties of Gaussian states, $\mathbf{Z}$ matrix and entanglement entropy}
%%%%%%%%%%%%%%%%%%%%%%%%%%%%%%%%%%
\label{appZ}

Both qubit cluster states and the qubit surface code can be represented by graphs. Strictly speaking, however, their definitions in terms of graphs are incompatible. In particular, qubit cluster states (or qubit ``graph states,'' as they are often confusingly called in the literature) are represented by graphs that act as a well-defined \emph{recipe} for creating the state: nodes represent qubits prepared in $|+\9$, and edges represent CPHASE gates between them. While one does not have to make cluster states this way, any cluster state can be so constructed. If we try to apply this recipe to the graphs for qubit surface-code states, however, we fail because for the surface code, qubits live on edges, while plaquettes and vertices define stabilizers in terms of all $\sz$'s or all $\sx$'s. There is no analog for this in the graph recipe used to define qubit cluster states. While one can \emph{define} by fiat a connection between the two types of graphs, it is at best a patch-up job.

In contrast, the graphical calculus for Gaussian pure states~\mboxcite{Menicucci2011} provides a \emph{unified} graphical representation of any Gaussian pure state in terms of a recipe for its creation. Again, one does not have to follow the recipe to make a Gaussian state, but any Gaussian state can be made by the recipe defined (uniquely) by its graph. As such, for CV cluster states and the CV surface code, only one type of graph is needed. Furthermore, the measurement patterns that connect the two types of states have definite graph transformation rules, so the connection between the two graphs can be \emph{derived} using the graphical calculus instead of invented ad hoc as a patch between two inequivalent types of graphs, as in the qubit case. This elegant connection is described in detail in this Appendix, where we also review the main features of Gaussian states.

Entanglement entropy of a generic state can be rather hard to compute since it generally requires knowledge of the spectrum of the density matrix. On the other hand, a zero-mean, $N$-mode Gaussian state can be described conveniently and completely by an easy algebraic formalism that follows from the form of its characteristic function \mboxcite{RP2011}, which is solely a function of the vector of the first statistical moments and the matrix $\Gamma$ that carries the information about the second moments. Recall that $\Gamma$ is the covariance matrix of the Gaussian state defined as
\beq
\Gamma_{j,k} = \text{Re} \tr [\rho \, \hat{\bar{r}}_i \hat{\bar{r}}_j],
\eeq
where $\hat{\bar{r}} = (\hat{q}_1, ..., \hat{q}_n, \hat{p}_1, ..., \hat{p}_n)^T$ is the $2N$-dimensional column vector of the hermitian quadrature operators of the $N$ modes. The information contained in the covariance matrix completely determines the entanglement properties of a Gaussian state \mboxcite{TD2012}. Explicit calculations of Gaussian states entanglement entropy are performed making use of the \textit{symplectic spectrum} of $\Gamma$. Let us introduce the symplectic form $\Omega$,
\beq
\Omega_{j,k} = -i [ \hat{\bar{r}}_i, \hat{\bar{r}}_j ] \, ,
\eeq
which is a skew-symmetric matrix that incapsulates the canonical commutation relations of the quadrature operators. For a Gaussian state $\rho$ with covariance matrix $\Gamma$, the positive elements of the $N$ pairs of eigenvalues $\{ \pm \sigma_i \}$ of the matrix product $i \Gamma \Omega$ are called \textit{symplectic eigenvalues}. 
%The important part of this construction is that the spectrum of $\Gamma \Omega$ is invariant under any symplectic transformation $S \in \text{Sp}(2n, \R)$ and thus characterizes the Gaussian state. 
The connection between entanglement entropy $S(\rho)$ and symplectic eigenvalues of $\rho$ is given by the formula
\begin{multline}
\label{formulase}
S(\rho) = \sum_{\{ \sigma_i \}} \left [  \left ( \sigma_i + \frac{1}{2} \right) \log_2  \left ( \sigma_i + \frac{1}{2} \right) \right. \\
- \left. \left ( \sigma_i - \frac{1}{2} \right) \log_2 \left ( \sigma_i - \frac{1}{2}  \right) \right] \, .
\end{multline}
The Gaussianity of the state is preserved under Gaussian unitary transformations $\hat{U}_G$. Each of these operations has a correspondent matrix representation $\mathbf{Y}$ that belongs to the Symplectic group Sp$(2n, \R)$, while quadratures measurements have a well-defined action on the covariance matrix $\Gamma$ of the state, which is described below.\\
A symplectic transformation $\mathbf{Y}$ preserves the canonical commutation relations as follows
\beq
\mathbf{Y} \Omega \mathbf{Y}^T = \Omega \, , \,\,\,\,\,\, \forall \, \mathbf{Y} \in  Sp(2n, \R) \, ,
\eeq 
while the action of a Gaussian transformation $\hat{U}_{G}$ on the quadratures can be expressed by
\beq
\hat{\bar{r}}^\prime = \hat{U}_G^\dagger \hat{\bar{r}} \hat{U}_G \longrightarrow \hat{\bar{r}}^\prime = \mathbf{Y} \hat{\bar{r}} \, ,
\eeq
where the right-hand side corresponds to a matrix multiplication on the quadratures vector. At the level of the covariance matrix, this is reflected in the transformation rule
\begin{align}
\Gamma^\prime = \text{Re} \6 \hat{\bar{r}}^\prime \hat{\bar{r}}^{\prime\, T} \9 \longrightarrow \Gamma^\prime &=  \text{Re} \6 \mathbf{Y} \hat{\bar{r}} ( \mathbf{Y} \hat{\bar{r}})^T \9 \nonumber \\ &= \mathbf{Y} \,\text{Re} \6 \hat{\bar{r}} \hat{\bar{r}}^T \9 \mathbf{Y}^T \nonumber \\
& = \mathbf{Y} \Gamma \mathbf{Y}^T \, .
\end{align}
A state of $N$ independent vacua is described by the covariance matrix 
\beq
\Gamma_0 = \frac{1}{2} \mathbf{I}_{2N} \, .
\eeq
\begin{figure}[t]
\includegraphics[width=0.8\columnwidth]{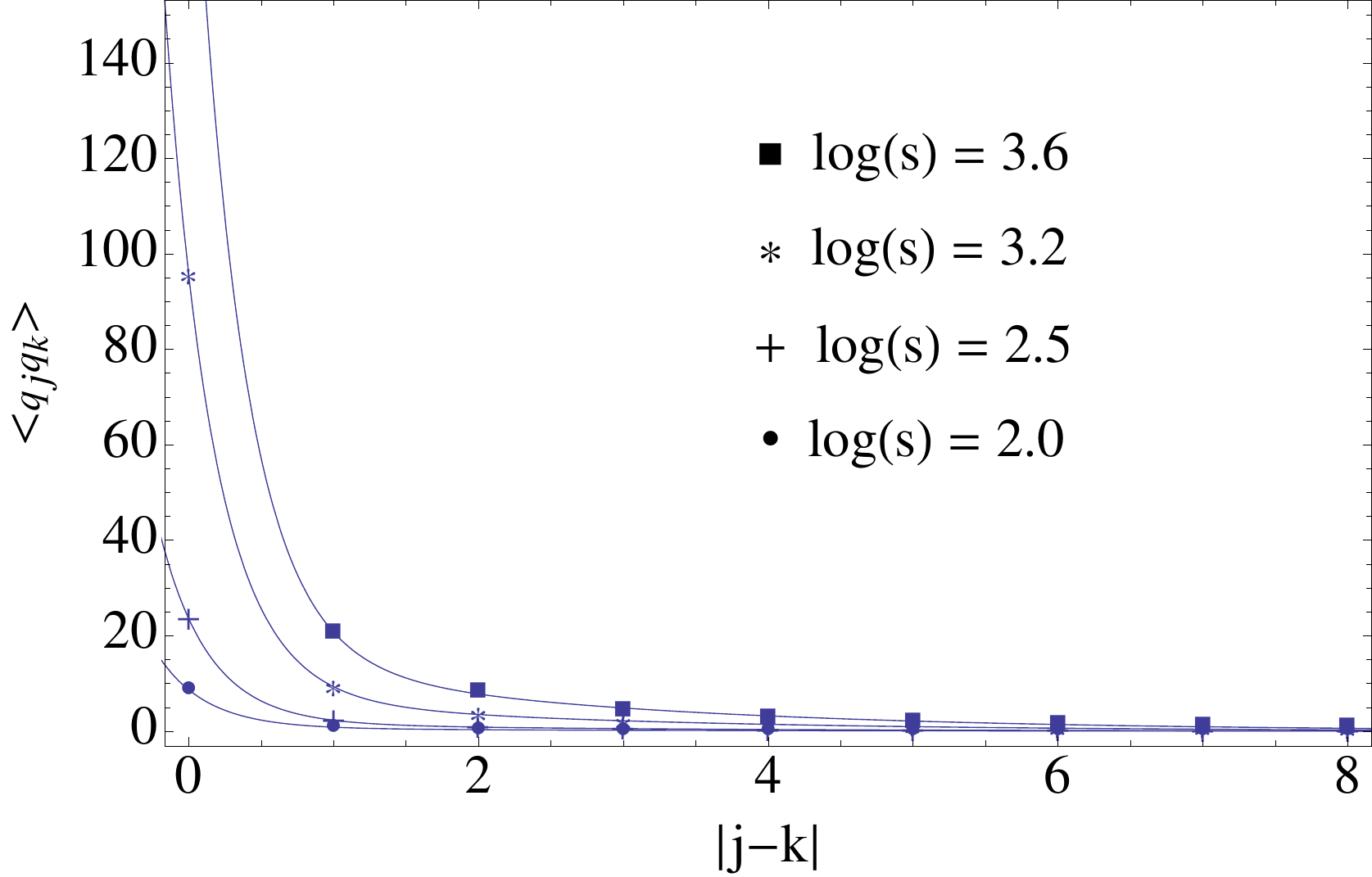}
\caption{$\hq$ correlations for the qumodes along the main axes of the open boundary CV surface code for different values $r$ of the initial squeezing. They decay as $a e^{-\frac{|j-k|}{\xi_a}} + b e^{-\frac{|j-k|}{\xi_b}}$ and for $\log(s)>2$ the correlation length scales converge quickly to $\xi_a=0.33,\xi_b=2.42$.  In contrast, $\hp$ correlations immediately drop to zero beyond one unit separation.  }
\label{corre}
\end{figure} 
After a Gaussian unitary transformation $\hat{U}_{\mathbf{Y}}$ represented by the matrix $\mathbf{Y}$ is applied to $\Gamma_0$, the resulting Gaussian state is
\begin{align}
\Gamma_{\mathbf{Y}} = \frac{1}{2} \mathbf{Y}\mathbf{Y}^T \, .
\end{align}
Exploiting the decomposition properties of symplectic matrices \mboxcite{Simon88,Menicucci2011}, the product $\mathbf{Y}\mathbf{Y}^T$ is uniquely specified by $\mathbf{Y}_{(\mathbf{U}\mathbf{V})}\mathbf{Y}^T_{(\mathbf{U}\mathbf{V})}$, with
\begin{align}
\label{UVdecomp}
\mathbf{Y}_{(\mathbf{U}\mathbf{V})} =  \left (
\begin{tabular}{c c}
$\mathbf{U}^{-1/2}$ & $0$\\
$\mathbf{V} \mathbf{U} ^{-1/2}$ & $\mathbf{U}^{1/2}$\\
\end{tabular} \right) \, ,
\end{align}
where, for an $N$-mode state, both $\mathbf{U}$ and $\mathbf{V}$ are $N \times N$ symmetric matrices, and $\mathbf{U} > 0$. Hence, the complex linear combination
\beq
\label{zmatrix}
\mathbf{Z} : = \mathbf{V} + i \mathbf{U}\, ,
\eeq
offers an alternative description for a pure Gaussian state. The graph~$\mathbf Z$ shows up directly in the position-space wavefunction~$\psi_{\mathbf Z}(\mathbf q)$ for an $N$-mode Gaussian state~$\ket {\psi_{\mathbf Z}}$:
\begin{align}
\label{eq:psifromZ}
	\psi_{\mathbf Z}(\mathbf q) &= \pi^{-N/4} (\det \mathbf U)^{1/4} \exp \left(\frac i 2 \mathbf q^T \mathbf Z \mathbf q \right)\,,
\end{align}
where $\mathbf q = (q_1, \dotsc, q_2)^T$~is a column vector of c-number position variables. For this state, $\Gamma_{\mathbf{Y}}$ from Eq.~\eqref{UVdecomp} can be rewritten as
\begin{align}
\label{gammay}
\Gamma_{\mathbf{Y}} = \frac{1}{2} \left (
\begin{tabular}{c c}
$\mathbf{U}^{-1}$ & $\mathbf{U}^{-1}\mathbf{V}$\\
$\mathbf{V} \mathbf{U} ^{-1}$ & $\mathbf{U} + \mathbf{V}\mathbf{U}^{-1}\mathbf{V}$\\
\end{tabular} \right) \, .
\end{align}
The matrix $\mathbf{Z}$ has a simple transformation rule under a symplectic transformation $\mathbf{Y}$. If $\mathbf{Y}$ is decomposed into block form
\begin{align}
\mathbf{Y} = \left(
\begin{tabular}{c c}
$\mathbf{A}$ & $\mathbf{B}$\\
$\mathbf{C}$ & $\mathbf{D}$\\
\end{tabular} \right) \, ,
\end{align}
then the $\mathbf{Z}^\prime$ matrix associated to the transformed state is given by
\beq
\mathbf{Z}^\prime = (\mathbf{C} + \mathbf{D} \mathbf{Z})(\mathbf{A}+\mathbf{B}\mathbf{Z})^{-1}\, .
\eeq
The usefulness of this approach lies in the simple transformation rules of $\mathbf{Z}$ for the most common laboratory procedures corresponding to Gaussian unitary transformations, as listed in \mboxcite{Menicucci2011}. This permits the study of Gaussian state evolution simply in terms of appropriate transformations on $\mathbf{Z}$. Furthermore, it is possible to define the $\mathbf{Z}_{CS}$ matrix for the CV cluster state directly, solely making use of the adjacency matrix $\mathbf{A_d}$ that describes the square-lattice pattern of connections among the modes. Explicitly:
\beq
\mathbf{Z}_{CS}(s) : = \mathbf{A_d} + i s^{-2} \mathbf{I}_{N} \, ,
\eeq
with squeezing parameter $\log{s}$ and $\mathbf{I}_{N}$ the $N \times N$ unit matrix. This is illustrated in Fig.~\ref{clusterandsurface}a. 
For the finitely squeezed cluster state with open boundary conditions, we start with $\mathbf{Z}_{CS}(s)$ and perform the measurement scheme displayed in Fig.~\ref{scheme1} in the main text. Measurements have a straightforward translation in the $\mathbf{Z}$ transformation rules language. A $\hq$~measurement on the $k$th mode is equivalent to deleting the $k$th row and column of the $\mathbf{Z}$ matrix, while a $\hp$ measurement is equivalent to applying a $\pi/2$ phase shift on the $k$th mode and then measuring the $\hq$ quadrature. At the level of the $\mathbf{Z}$ matrix, any $\hp$ measurement deletes the measured mode while generating new connections among its nearest neighbors.

\begin{figure}
\begin{tabular}{ccc}
\includegraphics[width=0.48\columnwidth]{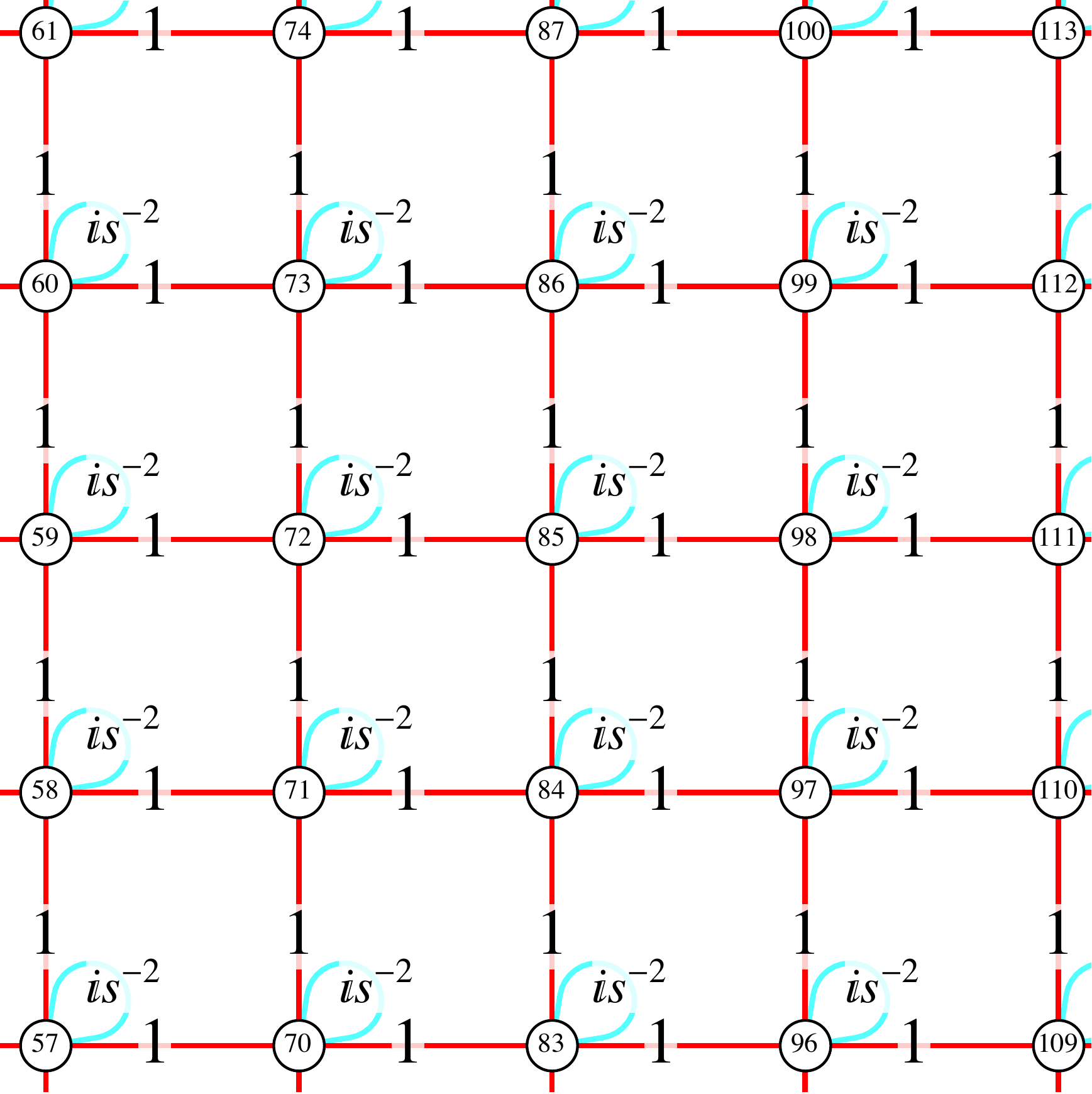} &  &
\includegraphics[width=0.48\columnwidth]{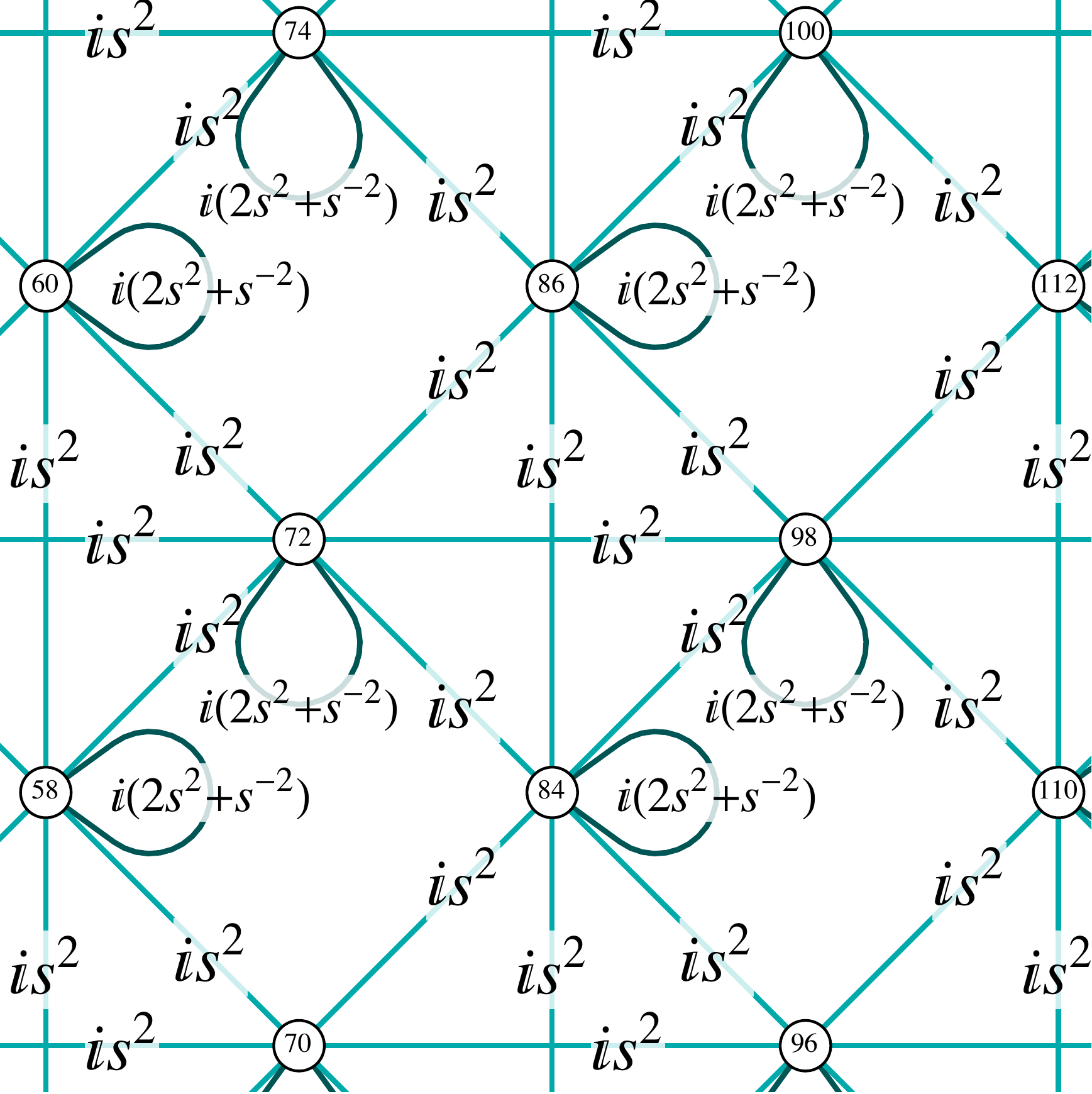}
 \\
\footnotesize{(a) CV cluster state} & & \footnotesize{(b) CV surface-code state}
\end{tabular}
\caption{Gaussian pure-state graph~$\mathbf Z$~\mboxcite{Menicucci2011} for a section of a canonical CV cluster state~(a) and for a section of the CV surface-code state~(b). Applying the measurement pattern prescribed in Fig.~\ref{scheme1}a in the main text to the state represented by the graph in~(a) produces a state with the graph in~(b). In both graphs, $\log s$ is called the squeezing parameter as used throughout this document. The darkness of each line is proportional to its magnitude, with color indicating phase (red = positive real, cyan = positive imaginary).
\label{clusterandsurface}}
\end{figure} 

The $\mathbf{Z}_{CS}(s)$ transforms into the CV finitely squeezed surface-code state $\mathbf{Z}_{\text{SC}}(s)$, shown in Fig.~\ref{clusterandsurface}b. In this case, it is a matrix having only imaginary entries given by
\begin{align}
\label{eq:ZSC}
	\mathbf{Z}_{\text{SC}}(s) &= i \mathbf{U}_{\text{SC}}(s)\,,
\end{align}
with
\begin{align}
\label{Usc}
\mathbf{U}_{\text{SC}}(s) &= s^2 \mathbf A_{\text{SC}} + \left( s^{-2} + 2 s^{2} \right ) \mathbf{I}_{N} \, \, ,
\end{align}
and $\mathbf A_{\text{SC}}$ adjacency matrix of the surface code. This immediately provides a simple expression for the covariance matrix of the newly evolved state:
\begin{align}
\label{cov}
\Gamma_{\text{SC}}(s) = \frac{1}{2} \left (
\begin{tabular}{c c}
$\mathbf{U}^{-1}_{\text{SC}}(s)$ & $0$\\
$0$ & $\mathbf{U}_{\text{SC}}(s)$\\
\end{tabular} \right) \, .
\end{align}
Complete knowledge of the covariance matrix makes straightforward calculations of entanglement entropy for different regions of the system. It is enough to derive the proper reduced covariance matrices, calculate the symplectic eigenvalues, and apply the formula from Eq.~\eqref{formulase}.

The covariance matrix $\Gamma_{\text{SC}}(s)$ allows us to study the $\hq$ and $\hp$-correlations for the modes of the surface code. From Eq. \eqref{Usc} one can immediately notice that the $\hp$-correlations are non-zero exclusively in the nearest neighborhood of each mode, due to the particular form of the matrix $\mathbf{U}_{\text{SC}}(s)$. The graph in Fig. \ref{corre} shows the behavior of the $\hq$-correlations along the main axes of the system described by Eq. \eqref{cov}, which are not simple because $\mathbf{U}^{-1}_{\text{SC}}(s)$ is complicated. These correlations decay exponentially with the distance.

%%%%%%%%%%%%%%%%%%%%%%%%%%%%%%%%%%
\section{Correlations on the lattice}
\label{latcorrs}
%%%%%%%%%%%%%%%%%%%%%%%%%%%%%%%%%%
A necessary assumption to use the topological entropy formulas is the exponential decay of the quadrature correlations. The covariance matrix $\Gamma_{\text{SC}}(s)$ allows us to study the $\hq$ and $\hp$-correlations for the modes of the surface code. From Eq. \eqref{Usc} one can immediately notice that the $\hp$-correlations are non-zero exclusively in the nearest neighborhood of each mode, due to the particular form of the matrix $\mathbf{U}_{\text{SC}}(s)$. This corresponds to a correlation length $\xi_p = 1$. On the other hand, the $\hq$-correlations require a more elaborate analysis because the matrix $\mathbf{U}^{-1}_{\text{SC}}(s)$ is more complicated.

It is however possible to prove analytically that also the $\hq$-correlations decay quickly enough. First we show that the spectral range of $\mathbf{U}_{\text{SC}}(s)$, denoted $\sigma(\mathbf{U}_{\text{SC}}(s))$, satisfies $\sigma(\mathbf{U}_{\text{SC}}(s)) \subset [ a, b]$ where $a=s^{-2}$ and $b=s^2(8 + s^{-4})$. For the minimum eigenvalue, note that for finite squeezing the matrix $\mathbf{U}_{\text{SC}}(s)/s^2$ is positive definite but for $s \to \infty$, some $\hq - \hq$ correlations become infinite. This indicates that the matrix is singular in that limit and the smallest eigenvalue is therefore zero. Adding finite squeezing shifts the spectrum of $\mathbf{U}_{\text{SC}}(s)/s^2$ by $s^{-4}$ so the minimum eigenvalue of $\mathbf{U}_{\text{SC}}(s)$ is $a= s^{-2}$. 

To derive the largest eigenvalue of $\mathbf{U}_{\text{SC}}(s)$, observe that for the surface code on a lattice with periodic boundaries, the $\mathbf{Z}$ graph associated with the adjacency matrix $\mathbf{A}_{\text{SC}}$ is regular with degree 6, i.e. each node connects to six others. The largest eigenvalue of the adjacency matrix of a regular graph is equal to the degree with an associated eigenvector $\nu = {1, ..., 1}$ \cite{CDS}, therefore the maximum eigenvalue of $\mathbf{A}_{\text{SC}}$ is 6. From these considerations it follows that the largest eigenvalue of $\mathbf{U}_{\text{SC}}(s)$ is $b = s^2 (8 + s^{-4})$. 
 For lattices with open boundary conditions, the eigenvalues of $\mathbf{A}_{\text{SC}}$ are upper bounded by the maximal degree with corrections that fall off with the system size, so the spectrum of $\mathbf{U}_{\text{SC}}(s)$ lies in the interval $[ a, b]$.
 
On a $n \times m$ lattice, the $nm \times nm$ matrix $\mathbf{U}_{\text{SC}}(s)$ is block tridiagonal with $n$ identical $m \times n$ matrices $A$ on the diagonal and identical $m \times m$ matrices $B$ on the immediate upper and lower blocks. Here the matrix coordinates $(i,j)$ correspond to Euclidean coordinates $((i_x, i_y),(j_x, j_y))$ on the lattice where $i = m i_x + i_y$ for $i_x \in \{ 0, ..., n-1 \}$ and $i_y \in \{ 0, ..., m-1 \}$, etc. It is convenient to define a \emph{graph distance} $d(i,j) = \text{max} \{ |i_x - j_x|, |i_y-j_x| \}$ between coordinates $(i_x, i_y)$ and $(j_x, j_y)$. Since away from the edges the ${\bf Z}$ graph is a union of square graph with a graph having two diagonal edges passing though every other plaquette, the graph distance is the number of edges on the shortest path between $(i_x,i_y)$ and $(j_x,j_y)$ and it satisfies $ed(i,j)/\sqrt{2}\leq d(i,j)\leq ed(i,j)$ where $ed(i,j)=\sqrt{(i_x-j_x)^2+(i_y-j_y)^2}$ is the Euclidean distance.    The matrix $A$ is itself tridiagonal with elements $\alpha=2s^2+s^{-2}$ on the main diagonal and $\beta=s^2$ on the immediate upper and lower diagonal.  The matrix $B$ is also tridiagonal with diagonal elements $\gamma=s^2$ and immediate upper and lower diagonal elements either equal to zero or $\gamma$.  A theorem of  Demko, Moss, and Smith \cite{DMS,Molinari} shows that banded matrices of a certain class have inverses with matrix elements that decay exponentially with the distance from the diagonal.  Specifically they show for matrices $M$ of size $N\times N$ and spectral range $\sigma(M)\subset [a,b]$ with $a>0$ [Ref. \cite{DMS} Proposition 5.1]:
\begin{equation}
\sup\{|M^{-1}_{i,j}|:(i,j)\in D_n(M)\}\leq C_0 q^{n+1}
\label{decaybound}
\end{equation}
where the \emph{decay sets} are
\[
D_n(M)=(\{1,\ldots N\}\times \{1,\ldots N\})\setminus S_p(M)
\]
and the \emph{support sets} are
\[
S_p(M)=\bigcup_{k=0}^p\{(i,j):M^k_{i,j}\neq 0\}.
\]
Here $C_0=\frac{(1+\sqrt{b/a})^2}{2b}$ and $q=\frac{\sqrt{b/a}-1}{\sqrt{b/a}+1}<1$.

The matrices relevant to our problem are in this class.  The matrix power $\mathbf{U}^k_{\text{SC}}(s)$ is a banded block symmetric matrix with blocks of size $m\times m$ and block band width $2k+1$.   Furthermore, each such block is banded with band width $2k+1$.     Thus the support set $S_p(\mathbf{U}_{\text{SC}}(s))$ is the set of those matrix coordinates $(i,j)$ such that the graph distance $d(i,j)$, is no more that $2p+1$.   Similarly, the decay set is all matrix coordinates outside the support set.  The statement in Eq. \ref{decaybound} is that for nodes separated in graph distance $d(i,j)>2p+1$ with associated matrix coordinates $(i,j)$ the inverse matrix element $\mathbf{U}^{-1}_{\text{SC}}(s)_{i,j}$ falls off exponentially with graph distance as $q^{(d(i,j)+1)/2}$.  This implies that the $ \hat{q}$-correlations between nodes $i$ and $j$ separated in graph distance by $d(i,j)$ satisfy 
\begin{equation}
\langle \hat{q}_i \hat{q}_j\rangle\leq Ce^{-(d(i,j)+1)/\xi}
\end{equation}
where the constant is
\[
C=\frac{(1+\sqrt{8s^{4}+1})^2}{4(8s^2+s^{-2})}
\]
and the correlation length is
\begin{equation}
\xi=\frac{2}{\ln \Big[\frac{\sqrt{8s^4+1}+1}{\sqrt{8s^4+1}-1}\Big]}.
\end{equation}

%%%%%%%%%%%%%%%%%%%%%%%%%%%%%%%%%%
\section{Topological entanglement entropy}
%%%%%%%%%%%%%%%%%%%%%%%%%%%%%%%%%%
\label{appTEE}

Let us give a physical interpretation of the different formulas used to identify the TEE. Topological order is a phase of matter resulting from non-local long-range correlations of topological origin that extend along all the modes of such systems. For this reason, local parameters cannot detect it, and therefore more effort is required in order to identify it. The first hint regarding the nature of the TEE was given in \mboxcite{AH2005}, where considerations about the toric code led the authors to note that the entanglement entropy of a region of the system is proportional to the perimeter of that region minus a constant. This idea was then broadened, and interesting relationships among entanglement entropy, total quantum dimension of the model, and TEE were introduced. The topology of the surface where a topologically ordered model is defined determines the species of anyons that are supported by the model. If we assign to each anyonic species a quantum dimension~$d$, and we sum over all the quasiparticles species, we call the \emph{total quantum dimension} the quantity
\beq
\D = \sqrt{\sum_i d_i^2}\, .
\eeq
Both Kitaev and Preskill (KP)~\mboxcite{AK2006} and Levin and Wen~(LW)~\mboxcite{ML2006} had the intuition that for a topologically ordered phase the entanglement entropy of a region scales as
\beq
S_A = \alpha |\partial A| - \gamma + \epsilon\, ,
\eeq
where $\alpha \in \R$, $|\partial A|$ is the size of the boundary of $A$, $\epsilon$ is a contribution that goes to zero in the limit of $|\partial A| \to \infty$, 
and
\beq
\label{topdim}
\gamma = \log \D
\eeq 
is the TEE. KP focus on constructing a linear combination of entanglement entropies such that all the terms proportional to the length of the boundaries of the regions cancel out, and only the topological term remains. Specifically, this corresponds to:
\begin{align}
\label{topoKP}
\gamma  \equiv - ( S_A &+ S_B + S_C - S_{AB}- S_{BC}- S_{AC} + S_{ABC}) \, .
\end{align}
Once the regions are chosen to be reasonably bigger than the correlation length of the system, it is also possible to demonstrate that deformations of the boundaries of the regions and smooth deformations of the Hamiltonian do not change the value of $\gamma$. This means that $\gamma$ is a quantity that is both invariant (i.e., only determined by the underlying topology) and universal (i.e., local modifications of the Hamiltonian do not affect the global topological properties). 

LW moved along a different line of thought, considering partitions of the system with different topologies rather than more general regions as KP, see Fig.~\ref{KPfigure}. The partitions are constructed such that their pairwise differences are equal, and chosen to be large enough, so that short-range correlations decay inside the boundaries. In analogy to Eq.~\eqref{topoKP}, an explicit expression for the TEE can be derived:
\begin{align}
\label{topoLW}
\gamma  \equiv - \frac{1}{2}[( S_A - S_B) - (S_C - S_{D}) ] \, .
\end{align}
Again, if the system is not topologically ordered, and if the entanglement entropies only depend on terms proportional to the boundaries, then the difference in Eq.~\eqref{topoLW} is zero. On the other hand, if topological long-range correlations are present on the lattice, LW argue that non-local closed string operators with non-vanishing expectation value exist on the lattice. Operators that wind around the region $A$ must then contribute to a nonzero value of Eq.~\eqref{topoLW}. They indeed call this contribution TEE, which is a witness for topological order in the following sense: the value of $\gamma$ characterizes the global anyonic and entanglement properties of the topological state. When $\gamma =0$, then it follows from Eq.~\eqref{topdim} that $\D = 1$, which physically means that no anyons are supported by the system, and no long-range topological correlations contribute to the entanglement. 
The idea behind both the KP and LW derivations of TEE formulas is to suppress the non-topological correlations, extracting only the topological information. 

For mixed states, the appropriate signature of topological order is given by a quantity called topological mutual information (TMI), introduced in \mboxcite{Iblisdir}:
\beq
\gamma_{\text{MI}} = - \frac{1}{2} (I_A + I_B + I_C - I_{AB} - I_{BC} - I_{AC} + I_{ABC})\, ,
\eeq
using the same regions used by KP for the calculations of the TEE. Now the correlations are measured using (one half of) the mutual information $I_{A}$ between a region $A$ of a system and its complement $A_c$, i.e., $I_{A} = S_A + S_{A_c} -S_{A\cup A_c}$, which replaces the von Neumann entropy $S_A$ (up to a factor of~2).

In \mboxcite{JW2012} a modified version of the TMI was proposed, to avoid treacherous ambiguities in the recognition of topological order. In fact, certain mixed states (in particular mixture of degenerate ground states of gapped local Hamiltonians) can exhibit long-range non topological correlations and therefore affect the value of the TMI as it was defined in \mboxcite{AK2006, ML2006}. Together with a new formulation for the TMI, they also introduced lower and upper bounds that precisely determine the TMI whenever they coincide. Consider now the LW regions as defined in Fig.~\ref{KPfigure} and label $E=A\setminus B$, set difference of $A$ and $B$, $F_1=A\setminus C$, and $F_2 = D$ (from which follows $F=F_1 \cap F_2=C)$. Then the value of the TMI lies between these two bounds:
\begin{equation}
\label{bounds}
 \text{min}(I_{E,F}-I_{E,F_1}-I_{E,F_2},0)\leq \gamma \leq \text{max}(I_{E,F_1},I_{E,F_2}). \\
\end{equation}
The circles ($\bullet $) in Fig.~3 in the main text, are a plot of the lower bound in Eq.~\eqref{bounds} in the worst-case scenario of very high initial temperature ($\kappa \to \infty$); see Appendix~\ref{app:TMIbigk}. This value for the bound is the lowest possible bound when creating the surface-code state from a thermal cluster state. It therefore illustrates the maximum extent to which the TMI can sink below the TEE for any given value of the squeezing parameter for the particular construction procedure described in the main text.

%------------------------------------------------------------------------------------------
\subsubsection*{Upper bound on TEE}
\label{upboundTEE}
%------------------------------------------------------------------------------------------

Here we compute an upper bound on the TEE for the CV surface-code state by calculating the subsystem entropy of a simpler network of entangled modes.  To do this, we invoke a the calculation of subsystem entropy appropriate to stabilizer states, which 
have vanishing two-point correlation functions.
Consider stabilizer states that are quantum doubles of a finite group $G$ (such as the toric code with group $G=\mathbb{Z}_2$).  As shown in Ref. \mboxcite{AH2005}, the TEE is calculated by dividing the system into two subsystems $A,\, B$ and identifying the redundant gauge transformations defined on the boundary between the two regions. The entanglement entropy of subsystem $A$ is the logarithm of the number of the (all equivalent) Schmidt coefficients of the state. Exploiting the group properties of $G$ allows one to write the entropy as $S(\rho_A) = (|\partial A| - 1)\log_2{|G|} $, implying $\gamma=\log_2{|G|}=\log_2(\mathcal{D})$.

For the CV surface codes, it is complicated to extract an analogous exact expression for the entropy of a subsystem because the Schmidt coefficients are not equal as in the discrete case.  Furthermore, the TEE is infinite for infinitely squeezed CV surface code states, and the definition of quantum dimension is not so clear for finitely squeezed CV surface code states since we do not yet have a description of this model in terms of a quantum double of a group.    Nevertheless, we can go ahead and compute the subsystem entropy in the same way that would be done for the discrete case and treat this as a bound for the TEE of the CV surface code state. It is simply an upper bound because we are ignoring longer-range correlations that degrade the topological order, but since the correlation length is bounded for any finite amount of squeezing, this should be a reasonably tight bound.

\begin{figure}[t]
\includegraphics[width=\columnwidth]{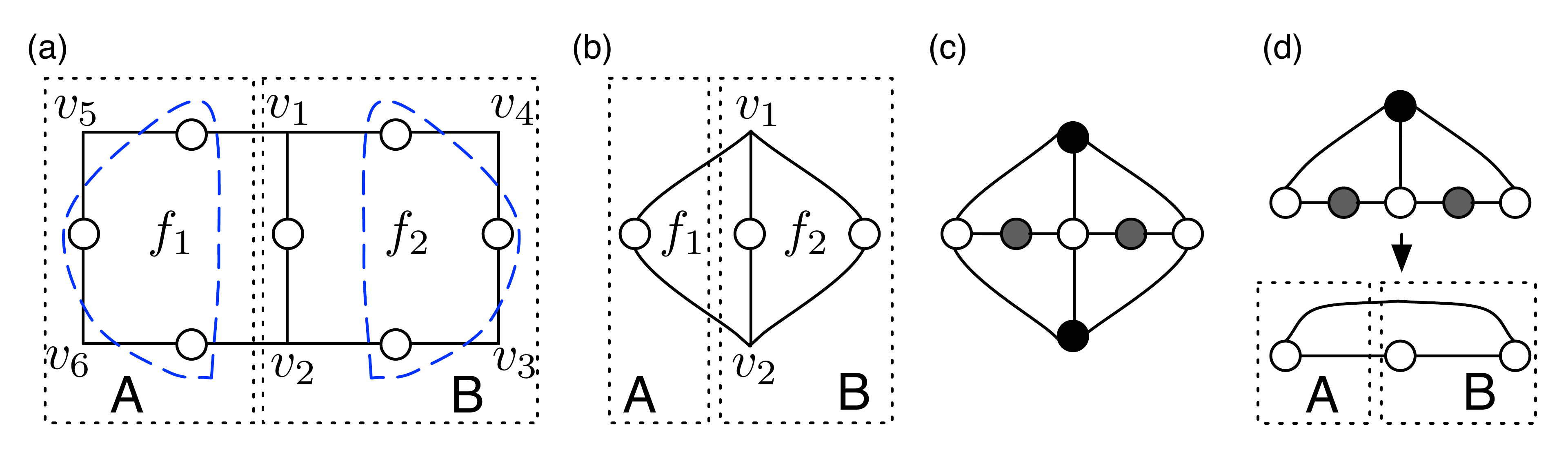}
\caption{Simplified network of modes with which to derive an upper bound on the TEE for the CV surface code:  (a)  This shows a surface-code state for a quantum double model with discrete variables.  There are six vertices, two faces, and seven edges where physical modes reside. Not all the vertex stabilizers are independent. Rather, their product is the identity, so the number of independent stabilizers is $5+2$, equal to the number of physical modes.  We can deform the lattice while preserving the topological order by replacing three modes of each face, excluding the mode on the shared boundary, with one mode as shown in (b).  This network has one independent vertex stabilizer, and two independent face stabilizers which equals the number of physical modes.  For the toric code, the network represents a ground state that is a GHZ state.  (c)  A finitely squeezed cluster-state graph with seven modes maps to the three-mode network by measuring the grey modes in the $\hq$ basis and the black modes in the $\hp$ basis.  In fact, the nullifiers on the top and bottom black modes of the graph act equivalently on the white modes meaning one of them is redundant. Therefore, an even simpler CV cluster-state graph suffices as depicted in (d).  Upon measurements, the reduced network has three modes with a correlation matrix that can be computed exactly to yield an upper bound to the CV surface code TEE.}
\label{ubounds}
\end{figure}

A simple configuration to start with is a quantum double model with a discrete group on a lattice with two faces.  This can be realised using a graph with just three edges (physical modes) and two vertices, as shown in Figs.~\ref{ubounds}a,b.  For the toric code the ground state would be the GHZ state $(|000\9+|111\9)/\sqrt{2}$ since both vertices implement the stabilizer $\hat{\sigma}^x_1\hat{\sigma}^x_2\hat{\sigma}^x_3$, and one face enforces $\hat{\sigma}^z_1\hat{\sigma}^z_2$, while the other face enforces $\hat{\sigma}^z_2\hat{\sigma}^z_3$.  Identifying two qubits on one of the faces with subsystem $B$, the subsystem entropy is $S(\rho_A)=S(\rho_B)=2-\gamma=1$, where $2$ comes form the size of the boundary of region $B$ and therefore $\gamma=1$.  This simplified surface-code network of three modes can be obtained from a finitely squeezed CV cluster state with six modes after measuring out three of the modes (Fig.~\ref{ubounds}d).   The resultant CV network has a correlation matrix that can be computed exactly. The symplectic spectrum for one subsystem consisting of one mode has two eigenvalues $\{\pm \sigma_1\}$ with
\begin{equation}
\sigma_{1} =\frac{1}{2}\sqrt{\frac{1 + 3 s^4+ 2 s^8}{1 + 3 s^4}}.
\end{equation}
Hence the TEE upper bound can be expressed as
\begin{equation}
\gamma^{\rm ub}(s)=  \Big [  \Big ( \sigma_1 + \frac{1}{2} \Big) \log_2  \Big ( \sigma_1+ \frac{1}{2} \Big) - \Big ( \sigma_1 - \frac{1}{2} \Big) \log_2 \Big ( \sigma_1 - \frac{1}{2}  \Big) \Big] \,.
\end{equation}
Asymptotically, the upper bound on TEE grows linearly with a slope of $\lim_{s\rightarrow\infty}\frac{d\gamma^{\rm ub}(s)}{d(\log s)} = 2/\ln(2)$.

%%%%%%%%%%%%%%%%%%%%%%%%%%%%%%%%%%
\section{Surface-code TMI from high-temperature CV cluster states}
%%%%%%%%%%%%%%%%%%%%%%%%%%%%%%%%%%
\label{app:TMIbigk}

In this appendix, we derive a bound for the topological mutual information (TMI), analyzing the limit $\kappa \to \infty$, which corresponds to the strongest possible decrease of the TMI from the TEE for the noise model considered. Recall that for a reduction $\rho_A$ of a pure state $\rho$, the von Neumann entropy $S(\rho_A)$ determines the entanglement entropy of the subsystem with respect to its complement. For Gaussian states, one can use the formula
\begin{multline}
\label{formula}
	S(\rho_A) = \sum_{i = 1}^{n_A^>} \left[ \left(\sigma_i + \frac{1}{2}\right)\log_2\left(\sigma_i + \frac{1}{2}\right) \right. \\
	\qquad - \left. \left(\sigma_i - \frac{1}{2}\right)\log_2\left(\sigma_i - \frac{1}{2}\right)   \right] \, ,
\end{multline}
where $\{ \sigma_i \}_A$ is the collection of $n_A^>$ symplectic eigenvalues associated to the reduced covariance matrix $\Gamma_A$ of the subsystem $\rho_A$. About the notation, in the following $n^{\ge}_X=n_X$ indicates the total number of symplectic eigenvalues $\ge \frac{1}{2}$ associated to a region $X$, $n^{>}_X$ indicates the number of eigenvalues $> \frac{1}{2}$, and $n^=_X$ denotes those $=\frac{1}{2}$.

Noise in our scheme has been modeled as beginning with a thermalized CV cluster state rather than a pure one. Since the normal-mode energies of the CV cluster-state Hamiltonian (Eq.~(\ref{sqCSHam}) in the main text) are all equal, this is equivalent to squeezing identical thermal states instead of vacuum states in the canonical construction procedure~\mboxcite{Menicucci2006}. The inverse temperature~$\beta$ of the state defines a useful parameter $\kappa= \coth({\beta /s^2})$.
To detect the topological order of the resulting mixed CV surface-code state, we use the TMI:
\beq
\label{eq:TMIapp}
\gamma_{MI} = - \frac{1}{2} (I_{A} + I_{B} + I_{C} - I_{AB} - I_{BC} - I_{AC} + I_{ABC}) \, ,
\eeq

The numerics show that the TMI does not decrease significantly with an increment of the initial value of $\kappa$. This is interesting and rather unintuitive, hence an analytical expression is required to confirm our findings. First of all, recall that the covariance matrix of the resulting mixed CV surface-code state is equivalent to the pure one ($\Gamma_0$) times $\kappa$, mathematically
\beq
\Gamma = \kappa \Gamma_0\, .
\eeq
As a consequence, the symplectic eigenvalues of $\Gamma$ (or any reduced section of it) are simply given by the "pure" symplectic eigenvalues multiplied by the overall $\kappa$ factor, $\{ \sigma_i \} = \{ \kappa \sigma_i^0 \}$. This simple transformation of $\Gamma$ is a special case that only arises due to the fact that all normal modes of the CV cluster-state Hamiltonian are identical, resulting in equal symplectic eigenvalues~$\kappa/2$.

\begin{figure}[t] %  figure placement: here, top, bottom, or page
   \centering
   \includegraphics[width=1.6in]{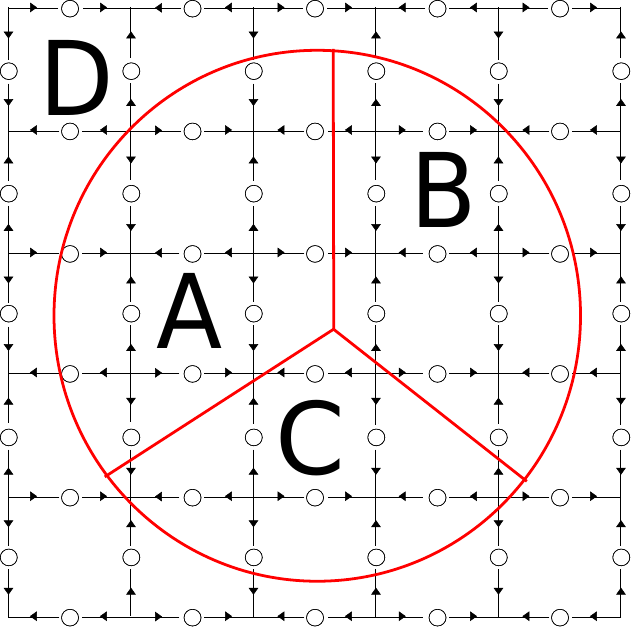} 
   \caption{Regions used for the TMI calculations}
   \label{areaklim}
\end{figure}

For a large value of $\kappa$ in Eq.~\eqref{formula}, we find the following asymptotic expression:
\begin{align}
\label{approxform}
\left(\kappa\sigma + \frac{1}{2}\right) \log_2\left(\kappa\sigma + \frac{1}{2}\right) & - \left(\kappa\sigma - \frac{1}{2}\right)\log_2\left(\kappa\sigma - \frac{1}{2}\right) \nonumber \\
&\approx \log_2(e\kappa \sigma) \, .
\end{align}
We can use this to show the behavior of the TMI as $\kappa \to \infty$. To start, consider the first term of Eq. \eqref{eq:TMIapp}: 
\beq
I_A = S_A + S_{BCD} - S_{ABCD} \, ,
\eeq
where the regions are shown in Fig. \ref{areaklim}. If the total state $A B C D$ has $N$ modes, then, for region A, we have $n_A$ modes and for its complement $n_{BCD}$ modes, such that $n_A + n_{BCD} = N$.
Hence, the von Neumann entropy for $A$ is given by
\begin{multline}
S_A =\sum_{i=1}^{n_A} \left[ \left(\kappa \sigma_i^A + \frac{1}{2}\right)\log_2\left(\kappa \sigma_i^A + \frac{1}{2}\right) \right. \\
	- \left. \left(\kappa \sigma_i^A - \frac{1}{2}\right)\log_2\left(\kappa \sigma_i^A - \frac{1}{2}\right)   \right] \, .
\end{multline}
In the limit of very high temperature, corresponding to ${\kappa \to \infty}$, we can use the approximation (\ref{approxform}) and rewrite the von Neumann entropy $S_A$ as
\begin{align}
\label{MIA}
\lim_{\kappa \to \infty} S_A \approx S_A^l &= \sum_{i=1}^{n_A} \log_2(e \kappa \sigma_i^A) %\nonumber \\ %= \sum_{i=1}^{n_A} \log_2(e \sigma_i^A) + \sum_{i=1}^{n_A} \log_2 \kappa =  \\
= \sum_{i=1}^{n_A} \log_2(e \sigma_i^A) + n_A \log_2 \kappa\, .
\end{align}
Divide now the $n_A$ symplectic eigenvalues into the two sets $n_A^>, n_A^=$, such that %. For the ones $> 1/2$, call the set $n_A^>$; for the ones equal to $1/2$, label the set $n_A^=$, with 
$n_A^> + n_A^= = n_A$. Hence, we can rewrite the expression in Eq. \eqref{MIA} as:
\beq
S_A^l =  \sum_{i=1}^{n_A^>} \log_2(e \sigma_i^A) + \sum_{i=1}^{n_A^=} \log_2 \left(\frac{e}{2} \right)  + n_A \log_2 \kappa \,.
\eeq
We can repeat the same argument for the region $BCD$ and find that
\begin{align}
S_{BCD}^l =  \sum_{i=1}^{n_{BCD}^>} \log_2(e \sigma_i^{BCD}) 
+ \sum_{i=1}^{n_{BCD}^=} \log_2 \left(\frac{e}{2} \right)  + n_{BCD} \log_2 \kappa \,.
\end{align}
The $N$ symplectic eigenvalues of $ABCD$ are all equal to $\kappa/2$. Consequently,
\beq
S_{ABCD}^l = N \log_2\kappa + N \log_2 \left(\frac{e}{2} \right) \, ,
\eeq
and the value for the mutual information $I_A^l$ in the $\kappa \to \infty$ limit is given by
\begin{align}
I_A^l &= S_A^l + S_{BCD}^l - S_{ABCD}^l \nonumber \\
&= \sum_{i=1}^{n_A^>} \log_2(e \sigma_i^A)
+\sum_{i=1}^{n_{BCD}^>} \log_2(e \sigma_i^{BCD}) \nonumber \\
&\qquad + (n_A^= + n_{BCD}^= - N) \log_2 \left(\frac{e}{2} \right) \, .
\end{align}
Notice that the $\kappa$-contributions cancel out exactly. Using $n_A^= + n_{BCD}^= = N - n_A^> - n_{BCD}^>$ and the area law behavior for the entropy, i.e., for regions sufficiently big, $n_A^> = n_{BCD}^>$ (although this does not mean that the sets $\{ \sigma_i^A \}$ and $\{ \sigma_i^{BCD} \}$ are the same), we can rewrite the mutual information as
\beq
I_A^l = \sum_{i=1}^{n_A^>} \log_2(e \sigma_i^A) + \sum_{i=1}^{n_{BCD}^>} \log_2(e \sigma_i^{BCD}) - 2 n_A^>  \log_2 \left(\frac{e}{2} \right)\,.
\eeq
The same argument applies for each other contribution to the TMI---for example, for region $C$:
\beq
I_C^l = \sum_{i=1}^{n_C^>} \log_2(e \sigma_i^C) + \sum_{i=1}^{n_{ABD}^>} \log_2(e \sigma_i^{ABD}) - 2 n_C^>  \log_2 \left(\frac{e}{2} \right)\,. 
\eeq
When substituting these expressions for the mutual information into the TMI formula in Eq.\eqref{eq:TMIapp}, all the elements $2 n_X^> \log_2(\frac{e}{2})$ sum to zero.

Consequently, the lower limit for the TMI is simply given by
\beq
\gamma_{MI}^{l} = -\frac{1}{2} \sum_{X} \signsymb(X) \sum_{i=1}^{n_X^>} \log_2 (e \sigma_i^X) \, ,
\eeq
where $X$ runs over all the possible combinations of regions, and the function $\signsymb(X)$ is defined as
\begin{align}
	\signsymb(X) &= 
	\begin{cases}
		+1 & \text{if $X \in \{A,B,C,D,ABC,ABD,ACD,BCD\}$} \\
		-1 & \text{if $X \in \{AB,AC,AD,BC,BD,CD\}$} 
	\end{cases}
	\nonumber \\
\end{align}
for a partitioning of the system as in Fig. \ref{areaklim}. 

%%%%%%%%%%%%%%%%%%%%%%%%%%%%%%%%%%

%%%%%%%%%%%%%%%%%%%%%%%%%%%%%%%%%%

%\bibliography{name_bibliography}

\end{document}